\documentclass[a4paper,10pt]{article}
\usepackage{stmaryrd}
\usepackage{amsfonts}
\usepackage{bbm}
\usepackage{amscd}
\usepackage{mathrsfs}
\usepackage{latexsym,amssymb,amsmath,amscd,amscd,amsthm,amsxtra,xypic}
\usepackage[dvips]{graphicx}
\usepackage[utf8]{inputenc}
\usepackage[T1]{fontenc}
\usepackage{lmodern}
\usepackage{amssymb}
\usepackage[all]{xy}
\usepackage{nicefrac,mathtools,enumitem}
\usepackage{microtype}
\usepackage{color}

\newtheorem{thm}{Theorem}[section]
\newtheorem{lem}[thm]{Lemma}
\newtheorem{cor}[thm]{Corollary}
\newtheorem{pro}[thm]{Proposition}
\newtheorem{ex}[thm]{Example}

\newtheorem{defi}[thm]{Definition}

\newtheorem{rmk}[thm]{Remark}
\newtheorem{remark}[thm]{Remark}

\newcommand {\emptycomment}[1]{} 

\newcommand{\be }{\begin{equation}}
\newcommand{\ee }{\end{equation}}

\newcommand{\pf}{\noindent{\bf Proof.}\ }

\newcommand{\g}{\mathfrak g}

\newcommand{\Z}{\mathbb Z}

\renewcommand{\k}{\mathbbm k}

\newcommand{\huaG}{\frak{g}}
\newcommand{\huag}{\frak{g}}

\newcommand{\huaV}{\mathcal{V}}
\newcommand{\huaW}{\mathcal{W}}

\newcommand{\huaH}{\frak{h}}

\newcommand{\huaK}{\frak{k}}

\newcommand{\frka}{\mathfrak a}

\newcommand{\frkd}{\mathfrak d}

\newcommand{\frkg}{\mathfrak g}
\newcommand{\frkh}{\mathfrak h}

\newcommand{\frkk}{\mathfrak k}

\newcommand{\frkr}{\mathfrak r}

\def\qed{\hfill ~\vrule height6pt width6pt depth0pt}

\newcommand{\pair}[1]{{\bf \langle\!\langle} #1 {\bf \rangle\!\rangle }}

\newcommand{\Id}{{\rm{Id}}}

\newcommand{\dM}{\mathrm{d}}

\newcommand{\Hom}{\mathrm{Hom}}

\newcommand{\gl}{\mathfrak {gl}}

\newcommand{\End}{\mathrm{End}}
\newcommand{\ad}{\mathrm{ad}}
\newcommand{\add}{\frka\frkd}

\newcommand{\sgn}{\mathrm{sgn}}
\newcommand{\Ksgn}{\mathrm{Ksgn}}

\begin{document}

\title{ {Lie 2-bialgebras\thanks
 {Research supported by NSFC
(10920161, 11101179, 11271202), SRFDP (200800550015, 20100061120096) and the
German Research Foundation (Deutsche Forschungsgemeinschaft (DFG))
through the Institutional Strategy of the University of G\"ottingen.
 }}}
\author{Chengming Bai$^1$, Yunhe Sheng$^2$  and Chenchang Zhu$^3$\\
$^1$Chern Institute of Mathematics and LPMC, Nankai University,\\
Tianjin 300071, China \quad Email: baicm@nankai.edu.cn \\
$^2$Department of Mathematics, Jilin University,
 \\Changchun 130012, Jilin, China\quad Email: shengyh@jlu.edu.cn\\
 $^3$Courant Research Centre ``Higher Order Structures'',\\
Georg-August-University G$\rm\ddot{o}$ttingen, Bunsenstrasse 3-5,
37073,\\ G$\rm\ddot{o}$ttingen, Germany\quad
Email: zhu@uni-math.gwdg.de }
\date{}
\maketitle



\begin{abstract}
In this paper, we study Lie 2-bialgebras, paying special attention
 to coboundary ones, with the help of the cohomology theory of $L_\infty$-algebras with
coefficients in $L_\infty$-modules. We construct examples of strict
Lie 2-bialgebras from left-symmetric algebras (also known as pre-Lie
algebras) and symplectic Lie algebras (also called quasi-Frobenius Lie
algebras).
\end{abstract}


\section{Introduction}

In this paper, we study the concept of a Lie 2-bialgebra with the
hope of providing a certain categorification of the concept of a Lie
bialgebra. We find a series of equations which can serve as
2-graded classical Yang-Baxter equations, and give various
solutions, thus naturally generating examples of Lie 2-bialgebras.
For the construction of solutions, we consider left-symmetric
algebras (also known as pre-Lie algebras) and symplectic Lie
algebras (also called quasi-Frobenius Lie algebras). The integration of a
Lie 2-bialgebra to a quasi-Poisson Lie 2-group is studied in detail
in \cite{CSX:quasi-P-2-groups} .

A Lie bialgebra \cite{D}
 is the Lie-theoretic case of a
bialgebra: it is a set with a Lie algebra structure and a Lie coalgebra
one  which are compatible. Lie bialgebras are the infinitesimal
objects of Poisson-Lie groups. Both Lie bialgebras and Poisson-Lie
groups are considered as semiclassical limits of quantum groups. The
study of quasi-triangular quantum groups involves the solutions of
the quantum Yang-Baxter equations. In the classical limit, the
solutions of the classical Yang-Baxter equations provide examples of
Lie bialgebras.

A categorification of Lie algebras is provided by 2-term
$L_\infty$-algebras (they are also called Lie 2-algebras)
\cite{baez:2algebras}. The concept of an $L_\infty$-algebra (sometimes
called a strongly homotopy (sh) Lie algebra)  was originally
introduced in \cite{stasheff:introductionSHLA,stasheff:shla} as a
model for ``Lie algebras that satisfy Jacobi identity up to all
higher homotopies''. The structure of a Lie 2-algebra also appears in
string theory, higher symplectic geometry
\cite{baez:classicalstring,baez:string}, and Courant algebroids
\cite{Roytenbergphdthesis,shengzhu1}. Thus, to give a model for the
categorification of Lie bialgebras,  it is natural to consider
a
pair of Lie 2-algebra structures on the dual 2-term complexes of
vector spaces
 with some higher compatibility conditions, namely,  a
``2-term $L_\infty$-bialgebra''.  But how does one allow homotopy in
a Lie bialgebra structure? A very nice method is given by Kravchenko
in \cite{olga} via higher derived brackets
\cite{akmanhigherderivedbracket,voro} and Kosmann-Schwarzbach's big
bracket \cite{YvetteBigbr}. Given a vector space $V$, we view the
bracket $l\in \wedge^2 V^* \otimes V$ and the cobracket $c\in V^*
\otimes \wedge^2 V$ as elements in $\wedge^\bullet(V\oplus V^*)$.
Then a Lie bialgebra structure on $V$ is equivalent to $\pair{l+c,
l+c}=0$, where  $\pair{\cdot,\cdot}$ is the big bracket defined by
extending the natural pairing between $V$ and $V^*$ via the graded
Leibniz rule:
\[\pair{v, u\wedge w} = \pair{v, u} \wedge w+(-1)^{|u||v|} u \wedge \pair{v, w}.
\] Using this idea, Kravchenko then generalizes the above to a $\Z$-graded vector space
$V_\bullet$ and defines an $L_\infty$-bialgebra. From an operadic
point of view, the minimal resolution defines a homotopy-version,
$P_\infty$, of an algebra $P$, if the corresponding operad (or
dioperad or PROP) $O_P$ of $P$ is Koszul. Lie bialgebras can
correspond either to a PROP or to a dioperad, and both of them are
proved to be Koszul in \cite{Gan,vallette:prop}.  Taking the minimal
resolution gives us a notion of an $L_\infty$-bialgebra which is
exactly the one defined by Kravchenko above.

However, in this setting, although a 2-term $L_\infty$-bialgebra
gives a Lie 2-algebra structure on $V$, it does not give a Lie
2-algebra structure on the dual, $V^*$. If one expects that a good
categorification of Lie bialgebras should consist of  Lie 2-algebra
structures on $V$ and $V^*$, along with some compatibility
conditions between them, then Kravchenko's $L_\infty$-bialgebra
needs to be modified. In \cite{CSX:new}, the authors applied a
simple shift to solve this problem.  From an operadic point of view,
such shifts have already appeared in \cite{Merkulov2} motivated from
an apparently different motivation in deformation quantization.

We adapt the shifting trick to our setting and give the definition
of a Lie 2-bialgebra
 (see Definition \ref{defi:lie-bi-general}). We
observe that a first example of a Lie 2-bialgebra is a  Lie
bialgebra  viewed as a Lie 2-bialgebra (see Remark
\ref{rk:lie-bialgebra}). Furthermore, in the strict case, we
describe the compatibility conditions between brackets and
cobrackets as a cocycle condition (Theorem \ref{thm:c-cocycle}).
For this,  we develop the cohomology theory of an $L_\infty$-algebra
$L$ with coefficients in representations  on  $k$-term complexes of
vector spaces (known as $L_\infty$-modules). When we restrict to the
adjoint representation, we recover the cohomology studied in
\cite{Penkava}. We give the adjoint representation of $L$ in terms
of the big bracket. We also introduce Manin triples in this general
framework. Here, we see the advantage of the language of the big
bracket: it makes concepts and calculations very elegant and
intrinsic. However, the usage of the big bracket also has the
disadvantage that, under some circumstances, it is not explicit
enough to give examples. So, we then focus on the strict case in
Section \ref{sec:strict-case} and explain the above concepts in
concrete formulas familiar to general algebraists. Associated to any
$k$-term complex of vector spaces $\huaV$, there is a natural
 differential graded Lie algebra $\gl(\huaV)$
\cite{lada-markl,shengzhu2}, which plays the same role as $\gl(V)$
for a vector space $V$ in the classical case. In the strict case, it
is enough to look at a certain strict Lie 2-algebra obtained
by applying truncation to $\gl(\huaV)$. This simplification makes it possible to write down
concrete formulas. This is the content of Section
\ref{sect:lie2-bi}.

The above leads to our  study of strict Lie 2-bialgebras in Section
\ref{sect:strict-lie2-bi}.  \emptycomment{In fact, this section in
the article is written in the usual algebraic language without
various grades and shifts, and it is self-contained. Thus, readers
who find the $L_\infty$-language confusing can simply skip Section
\ref{sect:lie2-bi} and directly go to Section
\ref{sect:strict-lie2-bi}.} Guided by the classical theory of Lie
bialgebras, we explore, in explicit terms, various higher
corresponding objects---matched pairs, Manin triples, standard Manin
triples---and their relations.   We first study standard Manin
triples of strict Lie 2-algebras, which is equivalent to strict Lie
2-bialgebras. Then, we study the conditions under which the direct
sum of two strict Lie 2-algebras with representations on each other
is a strict Lie 2-algebra (Theorem \ref{thm:directsum}). A pair of strict
Lie 2-algebras with representations on each other satisfying these
conditions is called a  matched pair. Given a strict Lie
$2$-algebra $\huaG=({\frak g}_0,{\frak g}_{-1}, \dM, [\cdot,\cdot])$
and a $2$-cocycle $(\delta_0,\delta_1)\in\Hom(\frak g_0,\frak g_{-1}\otimes \frak g_0\oplus \frak
g_0\otimes \frak g_{-1}) \oplus \Hom(\frak g_{-1}, \frak
g_{-1}\otimes \frak g_{-1})$, the 2-cocycle $(\delta_0,\delta_1)$ defines a semidirect product Lie
algebra structure on $\frkg_{-1}^*\oplus\frkg_0^*$. This further implies that $\huaG^*=(\frak g_{-1}^*,\frak g_0^*,\dM^*, [\cdot,\cdot]^*)$ is a
strict Lie $2$-algebra such that $(\huaG, \huaG^*; \ad^*, \add^*)$
is a matched pair (Theorem \ref{thm:delta-cocycle}). Thus, we see
that Manin triples of strict Lie 2-algebras $(\frkg\oplus
\frkg^*;\frkg,\frkg^*)$, matched pairs of strict Lie 2-algebras
$(\frkg, \frkg^*;\ad^*,\add^*)$ and strict Lie 2-bialgebras
$(\frkg;(\delta_0,\delta_1))$ describe the same object. Furthermore, in
Section \ref{sect:cybe}, we focus on the coboundary case,  i.e., we
require $(\delta_0,\delta_1)$ to be an exact $2$-cocycle. Due to the
abundant content of the corresponding cohomology theory, we find
that there are more generalized $r$-matrices than in the classical
case. We work out a set of 2-graded classical Yang-Baxter
equations (2-graded CYBE)
whose solutions provide examples of Lie
2-bialgebras.

In Section \ref{sect:ep}, we construct various (coboundary) strict
Lie 2-bialgebras via explicit solutions of 2-graded CYBE given by
left-symmetric algebras. Left-symmetric algebras (or pre-Lie
algebras, Vinberg algebras, and etc.) arose from the study of affine
manifolds and affine Lie groups, convex homogeneous cones and
deformations of associative algebras. They appeared in many fields
in mathematics and mathematical physics (see the survey article
\cite{leftsymm4} and the references therein). In particular, there
is a close relationship between left-symmetric algebras and
classical Yang-Baxter equations, which leads to one  regarding  the
former as the algebraic structures   behind the latter
\cite{Bai-Unifiedapproach}. We use the classification of low
dimensional left-symmetric algebras to give an explicit example of a
low dimensional Lie 2-bialgebra (Example \ref{ep:r2}).

 Furthermore, left-symmetric
algebras are also regarded as the underlying algebraic structures of
symplectic Lie algebras \cite{chu}, which coincides with
 Drinfeld's  observation of the correspondence between the invertible (skew-symmetric) classical $r$-matrices and the symplectic forms
 on Lie algebras \cite{D}. We then construct a general type of Lie
 2-bialgebras arising naturally from symplectic Lie algebras (Example
 \ref{ep:symp}). The naturality of the construction suggests that there may be some
 geometric meaning of such Lie 2-bialgebras. This is, however, still
 a mystery to us.

\emptycomment{
 Here we mention that just as Kravchenko's $L_\infty$-bialgebra
 could be obtained from PROP approach \cite{Gan,Merkulov1,Merkulov2}, i.e. the algebra associated
 to the minimal resolution $LieBi_\infty$ of the PROP $LieBi$ of a Lie
 bialgebra, our Lie 2-bialgebra could also be obtained through the
 PROP approach, see Remark \ref{rmk:prop}. }

Finally, since Lie bialgebras can be viewed as semiclassical limits
of quantum groups, a natural question to ask is whether there is
some relation possibly via quantization, between our
categorification and Khovanov-Lauda's recent categorification of
quantum groups \cite{khovanov-lauda:cat1}. At this very early stage,
as far as we can tell, the two sorts of categorification are rather
different. Any relation, if existing, will be nontrivial to
establish. Also, we do not claim our work is the final word in the
categorification of Lie bialgebras with respect to the above.
Instead, we regard it as something which opens a rather interesting
direction, along which we are currently working. \vspace{2mm}

\noindent {\bf Notations:} DGLA is short for differential graded Lie
algebra;
 $x,y,z$ are arbitrary elements in $\frkg_0$; $x^*,y^*,z^*$ are arbitrary elements in $\frkg_0^*$; $h,k,l$ are arbitrary elements in $\frkg_{-1}$ and
 $h^*,k^*,l^*$ are arbitrary elements in $\frkg_{-1}^*$; for a graded vector space $V= \sum_{n\in \mathbb Z} V_n $, $V[l]$ denotes the $l$-shifted graded vector space, namely
$V[l]_n=V_{l+n}$; $Sym(V)$ is the symmetric algebra of $V$.

\vspace{2mm}
 \noindent {\bf Acknowledgement:} We give our warmest
thanks to Damien Calaque, Zhuo Chen, Zhang-Ju Liu, Yvette
Kosmann-Schwarzbach, Rajan Mehta, Weiwei Pan  and Ping Xu for very
useful comments and discussions. Our paper is also partially
motivated by Mathieu Sti\'enon's talk in Poisson 2010.
Last but not least,  we are specially grateful to Prof. Ping Xu for
his various remarks and his encouragement to publish this
paper.

We give our special thanks to the referees, who gave us many helpful
suggestions and brought the reference \cite{Merkulov1} to our
attention.

\section{Lie $2$-bialgebras}\label{sect:lie2-bi}

\subsection{Lie 2-algebras via $L_\infty$-algebras}
Lie algebras can be categorified to Lie 2-algebras.  For a good
introduction on this subject see \cite{baez:2algebras}.

\begin{defi}
An {\bf  $L_\infty$-algebra} is a graded  vector space $L=L_0\oplus
L_{-1}\oplus\cdots$ equipped with a system $\{l_k|~1\leq k<\infty\}$
of linear maps $l_k:\wedge^kL\longrightarrow L$ with degree
$\deg(l_k)=2-k$, where the exterior powers are interpreted in the
graded sense and the following relation with Koszul sign ``{\rm Ksgn}''  is
satisfied for all $n\geq0$:
\begin{equation}\label{eq:higher-jacobi}
\sum_{i+j=n+1}(-1)^{i(j-1)}\sum_{\sigma}\sgn(\sigma)\Ksgn(\sigma)l_j(l_i(x_{\sigma(1)},\cdots,x_{\sigma(i)}),x_{\sigma(i+1)},\cdots,x_{\sigma(n)})=0.
\end{equation}
Here the summation is taken over all $(i,n-i)$-unshuffles with
$i\geq1$. A {\bf Lie 2-algebra} is a $2$-term $L_\infty$-algebra
$L=L_0 \oplus L_{-1}$.
\end{defi}

A 1-term $L_\infty$-algebra $L_0$ is a Lie algebra in the usual
sense. The only possible nonzero brackets of a 2-term
$L_\infty$-algebra (a Lie 2-algebra) are  $l_1$, $l_2$, and $l_3$.
The compatibility condition \eqref{eq:higher-jacobi} implies that
$l_1$ is a graded derivation with respect to $l_2$, and $l_3$
controls the obstruction of the Jacobi identity of $l_2$.   A {\bf
strict Lie $2$-algebra} is a Lie 2-algebra with $l_3=0$. This
specifically tells us that,  a strict Lie 2-algebra $\huaG$ is
simply a complex of
 vector spaces
$\frkg_{-1}\stackrel{l_1=\dM}{\longrightarrow}\frkg_0$ equipped with
a graded Lie bracket $l_2=[\cdot,\cdot]: \frkg_i\times
\frkg_j\rightarrow\frkg_{i+j}$, where $-1\leq i+j\leq 0$, such that
for all $x,y,z\in \frkg_0$ and $h,k\in \frkg_{-1}$, we have
\begin{equation}\label{eq:strict-lie2}\left\{\begin{array}{l}
~[x,y]=-[y,x], \quad [x,h]=-[h,x], \quad[h,k]=0, \quad
\dM([x,h])=[x,\dM h],\quad [\dM
h,k]=[h,\dM k], \\
 ~[[x,y],z]+[[y,z],x]+[[z,x],y]=0, \quad [[x,y],h]+[[y,h],x]+[[h,x],y]=0.
\end{array}\right.
\end{equation}
 The
notions of a strict Lie 2-algebra, a crossed module of Lie algebras,
and a 2-term DGLA are equivalent. Equation \eqref{eq:strict-lie2} implies
that there is a semidirect product Lie algebra structure on
$\frkg_0\oplus \frkg_{-1}$ with the Lie bracket defined as follows,
$$
[x+h,y+k]_{\frkg_0\oplus \frkg_{-1}}:=[x,y]+[x,k]+[h,y].
$$

An $L_\infty$-algebra $L$ gives a differential graded commutative
algebra (d.g.c.a) structure on the graded symmetric algebra
\[ Sym (L^*[-1])= \underbrace{\k}_{{\text{degree}\; 0}} \oplus
\underbrace{L^*_0}_{{\text{degree}\; 1}} \oplus \underbrace{\big[
\wedge^2 L^*_0 \oplus L^*_{-1}\big]}_{\text{degree} \;2} \oplus
\underbrace{ \big[ \wedge^3 L^*_0 \oplus L^*_0 \otimes L^*_{-1}
\oplus L^*_{-2}\big]}_{\text{degree} \;3} \oplus \dots,
\] whose  degree 1 differential $\delta$ is given by dualizing
$\{l_i|1\leq i<\infty\}$\footnote{This works for finite dimensional $L_i$'s,
 which is
our setting in this paper.}. The generalized Jacobi
identity \eqref{eq:higher-jacobi} is equivalent to $\delta^2=0$.
Then an $L_\infty$-morphism $f: L \to L'$ is given by a d.g.c.a.
morphism $Sym ((L')^*[-1]) \to Sym( L^*[-1])$. It is more general
than a ``strict morphism'', namely a morphism preserves all the
brackets strictly. More precisely,

\begin{defi}{\rm\cite{baez:2algebras}}\label{defi:morphism}
 Let $\huaG$ and $\huaG^\prime$ be  two strict Lie $2$-algebras. A {\bf strict homomorphism}
 $f$
 from $\huaG$ to $\huaG^\prime$ consists of
 linear maps $f_0:\frkg_0\longrightarrow \frkg_0^\prime$
  and $f_1:\frkg_{-1}\longrightarrow \frkg_{-1}^\prime$ commuting with the
  differential, i.e.,
  $
f_0\circ \dM=\dM^\prime\circ f_1,
  $
  such that
\begin{equation}\label{eqn:DGLA morphism c 1}\left\{\begin{array}{rll}
~[f_0(x),f_0(y)]^\prime-f_0[x,y]&=&0,\\
~[f_0(x),f_1(h)]^\prime-f_1[x,h]&=&0.\end{array}\right.
\end{equation}
\end{defi}

\subsection{$L_\infty$-modules and $L_\infty$-cohomology}
Now we recall the definition of an $L_\infty$-module
\cite{lada-markl}. Given a $k$-term complex of vector spaces $\huaV:
V_{-k+1}\stackrel{\partial}{\longrightarrow}\cdots
V_{-1}\stackrel{\partial}{\longrightarrow}V_0$, the endomorphisms
(not necessarily preserving the degree) form a DGLA $\gl(\huaV)$
with the graded commutator bracket and a differential inherited
from $\partial$. This plays the same role as $\gl(V)$ in the
classical case for a vector space $V$ (see \cite{lada-markl,
shengzhu2} for details). We say that $\huaV$ is  an
{\bf $L_\infty$-module} of an  $L_\infty$-algebra $L$ if there is an $L_\infty$-morphism $L \to \gl(\huaV)$, in which $\gl(\huaV)$ is considered as an $L_\infty$-algebra.

Another  equivalent definition of an
$L_\infty$-module
\footnote{The
  equivalence is supposed to be well-known. For a detailed proof, we
  refer to Dehling's master thesis
\cite{malte-thesis}.} of $L$ is given via a generalized
Chevalley-Eilenberg complex of $L$. That is, an $L_\infty$-module
structure on a graded vector space $\huaV$ is  given by a degree 1
differential $D$ on the graded vector space \[ (Sym (L^*[-1])
\otimes \huaV)_n = \oplus_k Sym(L^*[-1])_k \otimes V_{n-k}. \] The
second definition is subtly different from the first.  In the second
definition, $L$ acts on a graded vector space $\huaV$, whereas
in the first, $\huaV$ is a complex of vector spaces. However in the
second definition, $\huaV$ can be treated as a complex with
differential encoded in $D$. Thus the apparent discrepancy between
the two definitions does not affect our application. The second way
of defining $L_\infty$-modules has the advantage that we can view
$(Sym (L^*[-1]) \otimes \huaV, D)$ as the cochain complex $
C^\bullet (L, \huaV)$ of $L$ with coefficient in its
$L_\infty$-module $\huaV$. Then its cohomology group $H^\bullet (L,
\huaV)$ is defined to be the {\bf $L_\infty$-algebra
  cohomology} of $L$ with  coefficients in $\huaV$.  We denote the set of  $L_\infty$-modules
of $L$ by $Rep^\infty(L)$, and a typical element in $Rep^\infty(L)$
by  $(\huaV, D_\huaV) $.

Given $(\huaV, D_\huaV) $, $(\huaW, D_\huaW) \in Rep^\infty(L)$,
there is a degree 1 differential $D$ on $Sym(L^*[-1])\otimes (\huaV
\otimes \huaW)$ uniquely determined by
\[ D(\eta \otimes \xi)= D_\huaV(\eta) \otimes \xi + (-1)^{|\eta|} \eta
\otimes D_\huaW (\xi),
\]
for all $\eta \in Sym(L^*[-1])\otimes \huaV$ and  $\xi \in
Sym(L^*[-1])\otimes \huaW$.
 Similarly, one can take the
symmetric algebra $Sym (\huaV)$, the wedge product   $\Lambda(
\huaV)$, and the dual $\huaV^*$ of $L_\infty$-modules.

\subsection{$L_\infty$-bialgebras}\label{sect:l-infty-bi}

\noindent $\bullet $ Big bracket:

 Given  a graded vector space $V= \sum_{n\in \mathbb Z} V_n
$, it is well-known that there is a graded version of big bracket
$\pair{\cdot,\cdot}$ on $Sym(V^*[l]) \otimes Sym(V[k]) \cong
Sym(V^*[l]\oplus V[k]) \cong Sym(T^*[l+k]V[k])$ by extending the
usual pairing between $V^*$ and $V$ via a graded Leibniz
rule\footnote{Here we specially thank Yvette Kosmann-Schwarzbach,
Rajan Mehta and Dimitry Roytenberg for their help on the signs and
on the history of various brackets \cite{mehta}.}
\begin{equation} \label{eq:g-leibniz}\begin{split}
\pair{u, v\wedge w}= \pair{u, v}\wedge w + (-1)^{(|u|+l+k) |v|}
v\wedge \pair{u, w}, \pair {u, v}=-(-1)^{(|u|+k+l)(|v|+k+l)}
\pair{v, u},\end{split}
\end{equation} where $u \in Sym(V^*[l]\oplus V[k]) _{|u|}$ and $v \in
Sym(V^*[l]\oplus V[k]) _{|v|}$. The big bracket is in fact the
canonical graded Poisson bracket on $T^*[l+k]V[k]$. Thus, we have a
graded Jacobi identity:
\begin{equation}\label{eq:g-jacobi}
\pair{u, \pair{v, w}}=\pair{\pair{u, v},
w}+(-1)^{(|u|+k+l)(|v|+k+l)} \pair{v,
  \pair{u, w}}.
\end{equation}

\noindent $\bullet $ $L_\infty$-algebras via the big bracket:

 Given an
$L_\infty$-algebra $L$, the bracket $l_i$ can be viewed as a degree
2 element in $Sym(L^*[-1])\otimes L$, for example:
\[l_2 : \wedge^2 L \to L, \quad \leadsto \quad l_2 \in \wedge^2 L^* \otimes L \subset Sym^2(L^*[-1])\otimes L
. \] With various shifts, $l_i$ can be viewed as a degree $2-l-k$ element in
$(Sym(L^*[-1])\otimes L[k])[l]$.

\begin{lem} \label{lem:degree-t} Given an element $t\in Sym(V^*[l] \oplus
  V[k])$, the degree of the operator $\pair{t,\cdot}: Sym(V^*[l] \oplus
  V[k])\longrightarrow Sym(V^*[l] \oplus
  V[k]) $ is $|t|+k+l$.
\end{lem}
\pf The statement follows from straightforward calculations. \qed

\begin{lem} \label{lem:delta-big-bracket}
A series of degree $2-k$ elements $l_i\in (Sym^i(V^*[-1]) \otimes
V[k])$ with $i=1, 2, \dots$ on $V=V_0\oplus V_{-1}
\oplus \dots$ gives an
$L_\infty$-algebra structure
 if and only if $\pair{\sum_{i=1}^\infty l_i,
\sum_{i=1}^\infty l_i}=0$.
\end{lem}
\pf The proof depends on a simple observation: there is a degree 1
operator $\delta$ on $Sym (V^*[-1])$ given by
\[ Sym (V^*[-1]) \xrightarrow{\delta:=\pair{\sum_i l_i, \cdot}} Sym
(V^*[-1]). \]
By Lemma \ref{lem:degree-t},
        the degree of $\delta$ is 1.
The graded Leibniz rule \eqref{eq:g-leibniz} of $\pair{\cdot,\cdot}$
further implies that $\delta$ is a derivation. On the other hand, the
sum $\sum_{i} l_i^*$ extended by graded Leibniz rule gives rise to a
degree 1 derivation $\delta':  Sym(V^*[-1]) \to Sym (V^*[-1])$. The
set  $\{l_i|i=1,2,\cdots\}$ gives rise to $L_\infty$-brackets on $V$ if
and only if $\delta'^2=0$. It is clear that
$\delta|_{V^*[-1]}=\delta'|_{V^*[-1]}$. Since both of them are
derivations, we claim that $\delta=\delta'$.  Thus, we only need to show
that
$\delta^2=0$ if and only if  $\pair{\sum_{i=1}^\infty l_i,
\sum_{i=1}^\infty l_i}=0$. This is indicated by the following
calculation: by the graded Jacobi identity, we have
\[ \pair{\sum_i l_i, \pair{\sum_i l_i, u}} = \pair{\sum_i l_i, \sum_i l_i} u
+ (-1)^{1\cdot 1} \pair{\sum_i l_i, \pair{\sum_i l_i, u}}, \] which
implies that $2 \delta^2 u= \pair{\sum_i l_i, \sum_i l_i} u$.
 \qed

\vspace{3mm}

\noindent $\bullet $ Lie 2-bialgebras via $L_\infty$-bialgebras:

 A similar
theory holds for the Lie bialgebra setting. That is, an
$L_\infty$-bialgebra also corresponds to a d.g.c.a., but with the
differential $\delta=\pair{ \sum_{p=1}^\infty\sum_{q=1}^\infty  t_{pq},\cdot}$ coming from
more complicated data including brackets, $t_{p1} \in
(Sym^p(V^*[-1]) \otimes V[-1])$, cobrackets, $t_{1p} \in (V^*[-1]
\otimes Sym^p( V[-1]))$ and their relations $t_{pq} \in
(Sym^p(V^*[-1]) \otimes Sym^q (V[-1]))$ for $p, q \ge 2$. Here, with
the various degree shifts, the $t_{pq}$'s have degree 1. This is
equivalent to requiring their total degree (without shifting) to be
1 as in \cite{olga}. Nothing
stops us from shifting further to adapt the notion to our
application, which leads to the following definition.

\begin{defi}\label{defi:lie-bi-general}
Let $V:=V_0\oplus V_{-1} \oplus \dots $ be a  $\Z^{\leq 0}$-graded
vector space. An {\bf $L_\infty [l, k]$-bialgebra} structure on $V$ is
given by degree $3+l+k $ elements $t_{p, q}\in Sym^{p,
  q}(V^*[-1-l] \otimes V[-1-k])$, for $p, q \in
 \{1, 2, \dots, \}$,
such that $\pair{\sum_{p=1}^\infty\sum_{q=1}^\infty  t_{p,q}, \sum_{p=1}^\infty\sum_{q=1}^\infty t_{p,q} }=0$. An {\bf
$L_\infty[l,k]$-quasi-bialgebra} (resp.  {\bf quasi-$L_\infty[l,k]$-bialgebra})
further allows that the indices $p$ (resp. $q$)
could possibly be $ 0$.

A {\bf Lie $2$-bialgebra} $V=V_0\oplus V_{-1}$ is a $2$-term
$L_\infty[0,1]$-bialgebra. A Lie $2$-bialgebra is {\bf strict} if
$t_{13}=t_{22}=t_{31}=0$.
\end{defi}

\begin{rmk}\label{rmk:lie2bi}
Given a Lie $2$-bialgebra $(V, t_{pq})$, by the degree reason, the
only non-zero $t_{pq}$'s are $t_{11}, t_{21}, t_{31},$ $ t_{22},
t_{12}, t_{13}$. Therefore, $t_{11}, t_{21}, t_{31}$ can be
understood as brackets on $V$, and we denote them by $l_1, l_2,
l_3$; $t_{11}, t_{12}, t_{13}$ can be understood as cobrackets on
$V$, and we denote them by $c_1, c_2, c_3$; $t_{22}$ can be
understood as  the relation between brackets and cobrackets.
\end{rmk}

\emptycomment{
For our application, we need a $2$-term $L_\infty[0,1]$-bialgebra
with the ``strict relation'':
\begin{defi}\label{defi:lie2-bi-general}
Let $V:=V_0\oplus V_{-1}$ be a $ 2$-term $\Z^{\leq 0}$-graded vector
space. A Lie $2$-bialgebra structure on $V$ is given by degree $4$
elements $l_p\in Sym^{p,
  1}(V^*[-1] \otimes V[-2])$
 and $c_q \in Sym^{1, q}( V^*[-1] \otimes   V[-2]) $, for $p, q \in
 \{1, 2, 3\}$,
such that   $l_1=c_1$ and $\pair{\sum_{p,q=2}^3 (c_q+l_p) +l_1,
\sum_{p,q=2}^3
  (c_q+l_p)+l_1}=0$.
\end{defi}

 In fact, with the above interpretation, since
the Lie 2-bialgebra we define above does not contain the
$t_{22}$-term, it is {\em relation-wise strict} in the sense that
the brackets and cobrackets satisfy
strict relations.}


  The terminology of a Lie $n$-bialgebra is also used in
  \cite{Merkulov1,Merkulov2}. The meaning of $n$ therein is the degree of
 the shifts,  which is the same as one of our shifts, rather than the
 number of terms.  For this reason, we first
  clarify a possible confusion of terminology, i.e.,
  Merkulov's Lie $2$-bialgebra is different from what we call a Lie
  $2$-bialgebra. However, by a straightforward comparison of
  \cite[Proposition 1.5.1]{Merkulov2} to our Definition
  \ref{defi:lie-bi-general}, we see that his $Lie^1Bi_\infty$-algebra is the same as
  our $L_\infty[1, 0]$-bialgebra, with the  exception that he works with the $\Z$-graded
  version, and we work with the $\Z^{\leq0}$-graded version because
  Lada-Markl's $L_\infty$-module theory is in this
  setting.  Moreover, as observed by Kravchenko,  her $L_\infty$-bialgebra is the same as
  Merkulov's $LieBi_\infty$-algebra, which is the same as our
  $L_\infty[0,0]$-bialgebra. Furthermore, Chen-Stienon-Xu's weak Lie
  2-algebra \cite[Definition 2.6]{CSX:new} is
  our 2-term $L_\infty[0,1]$-bialgebra without the term $t_{22}$, but their quasi-Lie 2-bialgebra  is different from ours.  Nevertheless their strict Lie 2-bialgebra is the same as the one in our paper.

Here, we compare our degree convention to that in other
 works. In our Definition \ref{defi:lie-bi-general} of a Lie $2$-bialgebra, $V_{-1}^*$ is
  of degree $2$ and $V_0^*$ is of degree $1$. This is
  not the same as Kravchenko's convention on degrees in \cite{olga},
  where, for an $L_\infty$-algebra $V$, the total degree of an
  element $\xi_1\cdots\xi_p\alpha_1\cdots\alpha_q$ in $\wedge  ^pV^*\otimes\wedge  ^qV$ is defined to be
  $\sum_{i=1}^p|\xi_i|+\sum_{i=1}^q|\alpha_i|+p+q-2$.  Let us
  explain this in the $2$-term case. The $L_\infty$-coalgebra
  structure in Kravchenko's setting is given by maps $\gamma_p:V\longrightarrow\wedge ^pV$ of
  total degree $1$. In particular, $\gamma_1\in V_{-1}^*\otimes V_0$, $\gamma_2$ is an element in $V_0^*\otimes\wedge ^2V_0\oplus
V_{-1}^*\otimes V_0\wedge
  V_{-1}$. So what one obtains is not the usual Lie $2$-algebra structure on
  $V_0^*\stackrel{\gamma_1^*}{\longrightarrow}V_{-1}^*$. The degree shift trick allows us to adjust the map $\gamma_2$
  to obtain the usual Lie $2$-algebra structure on
  $V_0^*\stackrel{\gamma_1^*}{\longrightarrow}V_{-1}^*$. In
  terms of total degrees, we define the total degree of an
  element $\xi_1\cdots\xi_p\alpha_1\cdots\alpha_q$ in $\wedge  ^pV^*\otimes\wedge  ^qV$  to be
  $\sum_{i=1}^p|\xi_i|+\sum_{i=1}^q|\alpha_i|+p+2q-3$. Thus, the $t_{pq}$'s
  have  total degree $1$ in the non-shifted complex.

We would like to justify our terminologies of
  $L_\infty[l,k]$-quasi-bialgebra and quasi-$L_\infty[l,k]$-algebra. There are two sorts of possible twists
  in the world of Lie bialgebras: quasi-Lie bialgebras and Lie quasi-bialgebras \cite{YvetteBigbr}. Both of them  are special examples of
  $L_\infty[l,k]$-quasi-bialgebras, with $V=V_0$ and shifts $l=k=0$. A quasi-Lie bialgebra has the only nonzero
  $t_{p,q}$'s being $t_{2,1}$ (bracket), $t_{1,2} $ (cobracket), and
  $t_{3,0}$ (the twist); a Lie quasi-bialgebra has nonzero $t_{2,1}$ (bracket), $t_{1,2} $ (cobracket), and
  $t_{0,3}$ (the twist). Thus, if we allow the index $p$ or $ q$ to be
  $0$,  we obtain
 quasi versions of $L_\infty$-bialgebras. Notice that $L_\infty[0,0]$-quasi-bialgebra is already
 discussed in \cite[Sectiton 4.5]{olga} under the name $L_\infty$-quasi-bialgebra. What is special for us here, is
 that we allows various shifts.

 In the $2$-truncated
  version, a reasonable definition for a Lie $2$-quasi-bialgebra,  for
  example, can be obtained by adding an extra term $\theta\in V_0\otimes V_{-1}\otimes
  V_{-1}$ in Definition \ref{defi:lie-bi-general}. Note that $\theta$ is also a degree $4$ element in $Sym^{0, 3}( V^*[-1] \otimes
  V[-2])$.

A Lie $2$-algebra has a Jacobi identity which holds up to  homotopy.  However,  the Lie brackets $l_i$ are still strictly
graded-commutative. Roytenberg has a notion of a weak Lie $2$-algebra
\cite{Roytenberg}, where the graded-commutativity  is also weakened
up to homotopy.  One might desire a universal homotopic version of
this, that is, a weak $L_\infty$-algebra, and one might wonder about
a corresponding bialgebra version. In the commutative version, such
a weak $C_\infty$-operad is provided by an $E_\infty$-operad. A
trick which might give a universal method to provide symmetries
up-to-homotopy is to express an operad as an algebra of a colored
operad, then take the minimal resolution of this colored
operad\footnote{Private communication with Bruno Vallette and
  Malte Dehling.}. This is currently under investigation. Thus, we postpone the study of weak
$L_\infty$-bialgebras to later works.

\begin{lem}\label{lem:infty-bi-delta} A series of degree $3+l+k$ elements $t_{p,q} \in
  Sym^{p,q}(V^*[-1-l] \otimes V[-1-k]) $  with $p, q \in \{1,
2, \dots, \}$ on $V:=V_0\oplus V_{-1} \oplus \dots$ gives an
$L_\infty$-bialgebra structure if and only if
$\delta:=\pair{\sum_{p=1}^\infty\sum_{q=1}^\infty t_{p,q}, \cdot}$ defines a
d.g.c.a. structure on $Sym (V^*[-1-l]) \otimes Sym( V[-1-k])$.
\end{lem}
\pf Given such degree $3+l+k$ elements $t_{p,q}$ on $V$, by Lemma
\ref{lem:degree-t}, $\delta$ has
  degree $1=3+l+k-1-l-1-k$. The degree $3+l+k$ elements  $t_{p,q}$ on
  $V$ define an $L_\infty$-bialgebra structure on $V$ if and only if
  $\pair{\sum_{p=1}^\infty\sum_{q=1}^\infty t_{p,q}, \sum_{p=1}^\infty\sum_{q=1}^\infty t_{p,q} }=0$. Then the rest of the proof follows from the graded Jacobi
  identity \eqref{eq:g-jacobi} and the graded Leibniz rule
  \eqref{eq:g-leibniz}  using the same method of the proof of Lemma \ref{lem:delta-big-bracket}.
\qed

\emptycomment{
We also have a 2-truncated version,
\begin{lem} A series of degree $4$ elements $l_i\in (Sym^i(V^*[-1])
  \otimes V[-2])$ and $c_i \in  (V^*[-1] \otimes Sym^i( V[-2]))$  with $i=1,
2, 3$ on $V=V_0\oplus V_{-1}$ gives a Lie $2$-bialgebra structure if
and only if $c_1=l_1$ and  $\delta:=\pair{\sum_{i=2}^3 (l_i+
c_i)+l_1, \cdot}$ defines a d.g.c.a. structure on $Sym (V^*[-1])
\otimes Sym( V[-2])$.
\end{lem}
\pf By the degree reason, $l_i=c_i=0$ for $i>
  3$, and the rest is the same as the proof of Lemma \ref{lem:infty-bi-delta}
\qed
}

\begin{pro}  \label{prop:both} If $(V, l_1=c_1, l_2, l_3,  c_2, c_3, t_{22})$ is a Lie $2$-bialgebra,
 then both $(V, l_1, l_2, l_3)$ and $(V^*[1], c_1, c_2, c_3)$ are  Lie
 $2$-algebras.
Here, $\{l_p\}$ and $\{c_q\}$ are  brackets and cobrackets
respectively as defined in Remark \ref{rmk:lie2bi}.
\end{pro}
\pf  By degree reason, $\pair{\sum_{i=2}^3(l_i+ c_i)+l_1+t_{22},
\sum_{i=2}^3 (l_i+ c_i)+l_1+ t_{22}}=0$ is equivalent to
\begin{equation} \label{eq:l+c}
\left\{\begin{array}{l} \pair{l_1, l_1}=0, \quad \pair{l_1,
l_2}=0,\quad \pair{l_2, l_2} + 2\pair{l_3, l_1}=0, \quad \pair{l_2,
l_3}=0, \quad \pair{l_3,l_3}=0,
\\
 \pair{c_1, c_2}=0,\quad \pair{c_2, c_2} + 2\pair{c_3,
c_1}=0,\quad \pair{c_2, c_3}=0,\quad \pair{c_3,c_3}=0
    \\
\pair{l_2, c_2}+\pair{l_1, t_{22}} =0, \quad \pair{l_2, c_3}
+\pair{c_2, t_{22}}=0, \quad \pair{l_3, c_2}+\pair{l_2, t_{22}}=0,
\quad \pair{l_3,c_3}+\pair{t_{22},t_{22}}=0.
\end{array}\right.
\end{equation}
The first two lines of equations are equivalent to  $\pair{\sum_{i=1}^3
l_i, \sum_{i=1}^3
  l_i}=0$ and $\pair{\sum_{i=1}^3 c_i, \sum_{i=1}^3 c_i}=0$ respectively. Notice that $\sum_{i=1}^3 l_i \in Sym(V^*[-1]) \otimes V[-2]$
and  $\sum_{i=1}^3 c_i \in V^*[-1] \otimes Sym (V[-2]) = Sym
((V^*[1])^*[-1]) \otimes (V^*[1]) [-2]$. By Lemma
\ref{lem:delta-big-bracket}, $\{l_i\}$ and $\{c_i\}$ give $L_\infty$-algebra structures on
$L$ and $L^*[1]$ respectively. \qed\vspace{3mm}

Given an $L_\infty$-algebra $L$, it is easy to describe its adjoint
representation on any shift $L[k]$ via the big bracket:
\[ D_{ad}:=\pair{ \sum_{i=1}^\infty l_i, \cdot}: \quad Sym (L^*[-1]) \otimes L[k] \to Sym (L^*[-1]) \otimes L[k].
\] This extends to the symmetric algebras,
\begin{equation}\label{eq:td}
\tilde{D}_{ad}:=\pair{ \sum_{i=1}^\infty l_i,\cdot}: \quad Sym
(L^*[-1]) \otimes Sym(L[k]) \to Sym (L^*[-1]) \otimes Sym(L[k]).
\end{equation} To justify that $D_{ad}$ and $\tilde{D}_{ad}$ indeed define  representations, we need
the following lemma.

\begin{lem}
With the above notations, we have $D_{ad}^2=0$ and
$\tilde{D}_{ad}^2=0$.
\end{lem}
\pf Since $l_1,l_2,\cdots$  are $L_\infty$-brackets for $L$,   $\sum_{i=1}^\infty l_i$ is a degree $2-k$ element in
  $Sym(L^*[-1])\otimes L[k]$, and thus in $Sym(L^*[-1])\otimes Sym(
  L[k])$.
  Moreover, By Lemma
  \ref{lem:delta-big-bracket}, $\pair{\sum_{i=1}^\infty l_i, \sum_{i=1}^\infty l_i}=0$.
  Thus,
\[ \pair{\sum_{i=1}^\infty l_i, \pair{\sum_{i=1}^\infty l_i, u}} = \pair{ \sum_{i=1}^\infty l_i, \sum_{i=1}^\infty l_i} u
+ (-1)^{(2-k-1+k)^2} \pair{\sum_{i=1}^\infty l_i,
\pair{\sum_{i=1}^\infty l_i, u}},
\] which implies both $D_{ad}^2=0$ and $\tilde{D}_{ad}^2=0$.
\qed\vspace{3mm}

Now, in the strict case, we can describe the compatibility
conditions between brackets and cobrackets as a cocycle
condition. Here we make use of the condition $t_{22}=0$ in the definition of a strict Lie
2-bialgebra.

\begin{thm}\label{thm:c-cocycle} Strict Lie $2$-algebras $(V:=V_0\oplus V_{-1}, l_1, l_2)$ and $(V^*[1], c_1, c_2)$ form a Lie $2$-bialgebra if and only if $l_1=c_1$ as elements in
$V^* \otimes V$,  and  $\sum_{i=1}^2 c_i$ (or  $c_2$) is a $4$-cocycle
representing an element in $H^4(V, Sym( V[-2]))$.
\end{thm}
\pf Given a strict Lie 2-bialgebra $V$, the sum of the cobrackets
$\sum_{i=1}^2 c_i \in  Sym^1 (V^*[-1]) \otimes Sym(V[-2])$ being a
degree $4$ element, is  a 4-cochain in $C^4(V, Sym(V[-2]))$. In the
strict case, $l_3=c_3=t_{22}=0$. Thus, \eqref{eq:l+c} implies that
$\pair{l_1+l_2, c_1+c_2}=0$. Moreover, since $l_1=c_1$,
$\pair{l_1+l_2, c_1}=0$ automatically holds as long as $(V, l_1,
l_2)$ is a strict Lie 2-algebra.  Thus, $\tilde{D}_{ad} (\sum_{i=1}^2
c_i )=0$ and  $\tilde{D}_{ad} (c_2 )=0$. The converse direction can
 be proved similarly.  \qed\vspace{3mm}

\noindent $\bullet $ Manin triples:

As in the classical case, we have yet another description of
$L_\infty$-bialgebras via Manin $L_\infty[k]$-triples. When $k=0$,
we obtain \cite[Def. 32]{olga}.

\begin{defi}\label{defi:manin-triple}
A {\bf Manin $L_\infty[k]$-triple}  is a triple of $L_\infty$-algebras
$(\frkk, \g, \g')$ equipped with a nondegenerate graded symmetric
bilinear form $S(\cdot,\cdot )$, such that
\begin{enumerate}
\item $S(\cdot,\cdot)$ has degree $k$,  that is, there is an identification of $\g^*$ with
$\g'[k]$ via $S(\cdot,\cdot)$;
\item $\g$, $\g'$ are $L_\infty$-subalgebras of $\frkk$ such
  that $\frkk=\g \oplus \g'$ as graded vector spaces;
\item $\g$ and $\g'$ are isotropic with respect to
  $S(\cdot,\cdot)$;
\item the $n$-bracket   $\lambda_n$ of the $L_\infty$-algebra structure
  on $\frkk$ are invariant with respect to $S(\cdot,\cdot)$, i.e.,
\[S(\lambda_n(a_1, \dots, a_n), a_0)=(-1)^{|a_n| |a_0|}S(\lambda_n(a_1, \dots, a_{n-1}, a_0), a_n).
\]
\end{enumerate}
 \end{defi}

\emptycomment{ Then a 2-term  Manin $L_\infty[1]$-triple is a Manin
$L_\infty[1]$-triple $(\frkk, \g, \g')$ such that all three
$L_\infty$-algebras have only 2-terms. We will explain this in more
concrete terms in Section \ref{sect:manin-triple}. }

Then as expected, shifted Manin $L_\infty$-triples are related to shifted $L_\infty$-bialgebras:
\begin{thm} \label{thm:manin-triple}
The notions of  Manin
$L_\infty[k]$-triple and $L_\infty[0,-k]$-bialgebra are equivalent.
\end{thm}
\pf The proof is done by adding some careful counting of degrees to
  the proof of  \cite[Theorem 33]{olga}. We refer to \cite{malte-thesis}
  for this treatment.
\qed
\begin{remark}
We have a one-to-one correspondence between shifted Manin triples and
$L_\infty$-bialgebras with one-sided shifts. However, there is no
fundamental difference between the shift $l$ and the shift $k$. If
we were in a $\Z$-graded setting instead of $\Z^{\leq
  0}$-graded setting, they would be  dual to each other, that
is, the dual of an $L_\infty[l,0]$-bialgebra would be an
$L_\infty[0,-l]$-bialgebra.
\end{remark}

\subsection{Strict case}\label{sec:strict-case}

Now we explain the abstract definitions given in previous sections
with explicit formulas in the case of a strict Lie 2-algebra. This
is a preparation for the next section, wherein we address strict Lie
2-bialgebras in a more classical setting. This is not redundant
because with the concrete picture, we can address the non-symmetric
version,  which makes it better connected to the usual algebraic
discussion of Lie bialgebras. In this case, given a complex of vector
spaces, $\huaV:V_{-k+1} \xrightarrow{\partial} \dots
\xrightarrow{\partial} V_0 $, what is important is the strict
Lie 2-algebra of the 2-truncation of the endomorphism DGLA
$\gl(\huaV)$, because an $L_\infty$-morphism $L \to \gl(\huaV)$ can
only see this part. We denote the truncation by $\End(\huaV)$,
\begin{equation}\label{eqn:dgla of 2}
\End(\huaV):\End^{-1}(\huaV)\stackrel{\delta}{\longrightarrow}\End^0_\partial(\huaV),
\end{equation}
where $\End^{-1}(\huaV)=\{ E\in \bigoplus_{i=-k+2}^0\ \Hom(V_i,V_{i-1})|[E,E]_C=0\} $
with $[\cdot,\cdot]_C$ being the natural commutator, and $
\End^0_\partial(\huaV)=\{E\in\bigoplus_{i=-k+1}^0\End(V_i)|E\circ\partial=\partial\circ
E\}$.
\begin{defi}
A {\bf strict representation} of a strict Lie $2$-algebra $\huaG$ on
a $k$-term complex of vector spaces $\huaV$  is a
   strict homomorphism $\mu=(\mu_0,\mu_{1})$ from $\huaG$ to the strict Lie $2$-algebra
   $\End(\huaV)$. We denote a strict representation by
$(\huaV;\mu)$.
 \end{defi}



Let $(\huaV;\mu)$ be a $k$-term strict representation of $\huaG$. To
obtain the cohomology $H^\bullet (\huaG, \huaV)$, the space of
$p$-cochains is $C^p(\huaG,\huaV):= (Sym(\huaG^*[-1])\otimes
 \huaV)_p$.  The differential operator $D$ can be explicitly written
 as
$$
D=\widehat{\dM}+d_\mu+\widehat{\partial}: C^p(\huag, \huaV) \to
C^{p+1}(\huag, \huaV).
$$
We explain each term explicitly:
the operator $\widehat{\dM}:\Hom((\wedge^p\frkg_0)\wedge (Sym^q\frkg_{-1}),V_s)\xrightarrow{~}\Hom((\wedge  ^{p-1}\frkg_0)\wedge
(Sym  ^{q+1}\frkg_{-1}),V_s)$ is of degree $1$, and is induced
by $\dM$,
\begin{eqnarray*}
 && \widehat{\dM}(f)(x_1,\cdots,x_{p-1},h_1,h_2,\cdots,h_{q+1})\\
 &=&(-1)^{p}\big(f(x_1,\cdots,x_{p-1},\dM
  h_1,h_2,\cdots,h_{q+1})+c.p.(h_1,\cdots,h_{q+1})\big),
\end{eqnarray*}
where $f\in C^p(\huag, \huaV)$,
$x_i\in \frkg_0$ and $h_j\in \frkg_{-1}$.  The operator $\widehat{\partial}:\Hom((\wedge  ^p\frkg_0)\wedge (Sym
^q\frkg_{-1}),V_s)\longrightarrow\Hom((\wedge  ^p\frkg_0)\wedge
(Sym ^q\frkg_{-1}),V_{s+1})$ is of  degree $1$, and is induced
by $\partial$,
\begin{equation}
  \widehat{\partial}(f)=(-1)^{p+2q}\partial\circ f.
\end{equation}
Finally, the operator $d_\mu$ can be written as $d_\mu=(d_\mu^{(1,0)},d_\mu^{(0,1)})$, where
$d_\mu^{(1,0)}:\Hom((\wedge  ^p\frkg_0)\wedge (Sym
^q\frkg_{-1}),V_s)\longrightarrow \Hom((\wedge  ^{p+1}\frkg_0)\wedge
(Sym  ^q\frkg_{-1}),V_s)$ is given by
\begin{eqnarray*}
&& d_\mu^{(1,0)}(f)(x_1,\cdots,x_{p+1},h_1,\cdots,h_{q})\\
 &=&\sum_{i=1}^{p+1}(-1)^{i+1}\mu_0(x_i)f(x_1,\cdots,\widehat{x_i},\cdots,x_{p+1},
  h_1,\cdots,h_{q})\\
  &&+\sum_{i<j}(-1)^{i+j}f([x_i,x_j],x_1,\cdots,\widehat{x_i},\cdots,\widehat{x_j}\cdots,x_{p+1},
  h_1,\cdots,h_{q})\\
  &&+\sum_{i,j}(-1)^{i}f(x_1,\cdots,\widehat{x_i},\cdots,x_{p+1},
  h_1,\cdots,[x_i,h_j],\cdots,h_{q}),
  \end{eqnarray*}
  and $d_\mu^{(0,1)}:\Hom((\wedge  ^p\frkg_0)\wedge (Sym  ^q\frkg_{-1}),V_s)\longrightarrow
\Hom((\wedge  ^p\frkg_0)\wedge (Sym  ^{q+1}\frkg_{-1}),V_{s-1})$
is given by
\begin{eqnarray*}
 d_\mu^{(0,1)}(f)(x_1,\cdots,x_{p},h_1,\cdots,h_{q+1})
 &=&\sum_{i=1}^{q+1}(-1)^{p}\mu_1(h_i)f(x_1,\cdots,x_{p},
  h_1,\cdots,\widehat{h_i},\cdots,h_{q+1}).
\end{eqnarray*}

For any strict representation $(\huaV;\mu)$ of $\huaG$, let
$\huaV^*:V_0^*\stackrel{\partial^*}{\longrightarrow}V_{-1}^*\stackrel{\partial^*}{\longrightarrow}\cdots
V_{-k+1}^*$ be the dual complex of $\huaV$. The dual representation
$\mu_0^*:\frkg_0\longrightarrow\bigoplus_{i=-k+1}^0\End(V_i^*)$ and
$\mu_1^*:\frkg_{-1}\longrightarrow \End^{-1}(\huaV^*)$ can be defined by
\begin{eqnarray*}
  \langle \mu_0^*(x) u^*, v\rangle&=&-\langle u^*,
  \mu_0(x)v\rangle,\quad \forall~ u^*\in V_i^*,v\in V_i,\\
 \langle \mu_1^*(h) p^*, u\rangle&=&-\langle p^*,
\mu_1(h) u\rangle,\quad \forall ~ p^*\in V_{i}^*, u\in V_{i+1}.
\end{eqnarray*}
In fact, it is straightforward to see that $\mu_0^*$ commutes with
$\partial^*$, i.e.,  $\mu_0^*\in\End^0_{\partial^*}(\huaV^*)$.
Furthermore, $\mu^*\triangleq(\mu_0^*,\mu_1^*)$ is a strict
homomorphism from $\huaG$ to $\End(\huaV^*)$.


 If both
$(\huaV;\mu^V)$ and $(\huaW;\mu^W)$ are strict representations of
$\huaG$, then the tensor product $(\huaV\otimes\huaW;\mu)$ is also a
strict representation of $\huaG$, where $\mu=(\mu_0,\mu_1)$ is
explicitly given by
$$
\mu_0=\mu^V_0\otimes1+1\otimes\mu^W_0,\quad
\mu_1=\mu^V_1\otimes1+1\otimes\mu^W_1.
$$

The adjoint representation of $\huaG$ on itself, denoted by
$\ad=(\ad_0,\ad_1)$, with,
\begin{eqnarray*}
  \ad_0(x)=[x,\cdot]\in\End^0_\dM(\huaG),\quad
  \ad_1(h)=[h,\cdot]\in\End^1(\huaG),
\end{eqnarray*} is a strict
representation. The dual representation of $\huaG$ on $\huaG^*$ is
called the {\bf coadjoint representation} and denoted by
$\ad^*=(\ad_0^*,\ad_1^*)$. Then $\huaG$ acts on
$\huaG[-1]\otimes\huaG[-1]$---a 3-term complex of vector spaces
\begin{equation}\label{eq:gg}
(\huaG[-1]\otimes\huaG[-1])_{0}:=\frkg_{-1}\otimes\frkg_{-1}\stackrel{\dM^\otimes}{\longrightarrow}(\huaG[-1]\otimes\huaG[-1])_{1}
:=\frkg_{0}\otimes\frkg_{-1}\oplus\frkg_{-1}\otimes\frkg_{0}
\stackrel{\dM^\otimes}{\longrightarrow}(\huaG[-1]\otimes\huaG[-1])_2:=\frkg_{0}\otimes\frkg_{0},
\end{equation}
with $\dM^\otimes$  given by
\begin{eqnarray*}
  \dM^\otimes(h\otimes k)&=&(\dM\otimes1+1\otimes\dM)(h\otimes k)=\dM h\otimes k+h\otimes \dM k, \quad h,k\in \frak g_{-1},\\
  \dM^\otimes(x\otimes k+h\otimes y)&=&(\dM\otimes1-1\otimes\dM)(x\otimes k+h\otimes y)=\dM h\otimes
  y-x\otimes \dM k, \quad x,y\in \frak g_0, h,k\in \frak g_{-1}.
\end{eqnarray*}
 This
representation  plays an essential role in the next sections.
\emptycomment{when we consider the strict
Lie 2-bialgebras: thanks to the strict setting,  we do not have to
work with $Sym( \huaG)$ in its entirety as we have done in the
nonstrict case (see also Remark \ref{rk:degree} for the different
degree shift). }The corresponding Chevalley-Eilenberg complex  is
given by
\begin{eqnarray}
 &&\nonumber (\huaG[-1]\otimes\huaG[-1])_{0}\stackrel{D}{\longrightarrow} (\huaG[-1]\otimes\huaG[-1])_{1}\oplus
  \Hom(\frkg_{0},(\huaG[-1]\otimes\huaG[-1])_{0})\stackrel{D}{\longrightarrow}\\
&&\nonumber(\huaG[-1]\otimes\huaG[-1])_2
\oplus\Hom(\frkg_{0},(\huaG[-1]\otimes\huaG[-1])_{1})\oplus
  \Hom(\frkg_{-1},(\huaG[-1]\otimes\huaG[-1])_{0})\oplus\\
  &&\nonumber\Hom(\wedge  ^2\frkg_0,(\huaG[-1]\otimes\huaG[-1])_{0})\stackrel{D}{\longrightarrow}
  \Hom(\frkg_0,(\huaG[-1]\otimes\huaG[-1])_2)\oplus\Hom(\frkg_{-1},(\huaG[-1]\otimes\huaG[-1])_{1})\oplus\\&&\nonumber\Hom(\wedge  ^2\frkg_0,(\huaG[-1]\otimes\huaG[-1])_{1})
  \oplus\Hom(\wedge  ^3\frkg_0,(\huaG[-1]\otimes\huaG[-1])_{0})\oplus
  \Hom(\frkg_0\otimes
  \frkg_{-1},(\huaG[-1]\otimes\huaG[-1])_{0})\\\label{eq:gog}&&\stackrel{D}{\longrightarrow}\cdots,
\end{eqnarray}
where $D=\widehat{\dM}+d_\ad+\widehat{\dM^\otimes}$, in which
$d_\ad$ is the operator associated to the tensor representation
$(\ad_0\otimes 1+1\otimes\ad_0,\ad_1\otimes 1+1\otimes\ad_1)$ of
$\frkg$ on $\frkg[-1]\otimes\frkg[-1]$. For a 2-cochain
$(\delta_0,\delta_1)\in\Hom(\frkg_{0},(\huaG[-1]\otimes\huaG[-1])_{1})\oplus
  \Hom(\frkg_{-1},(\huaG[-1]\otimes\huaG[-1])_{0})$, we have
$
D(\delta_0,\delta_1)=-\dM^\otimes\circ\delta_0+d_\ad\delta_0-\delta_0\circ
\dM+\dM^\otimes\circ\delta_1+d_\ad\delta_1. $ Thus
$(\delta_0,\delta_1)$ is a $2$-cocycle if and only if the following
equations hold:
\begin{eqnarray}
  \label{eq:closed1}(\dM\otimes
  1-1\otimes\dM)\circ\delta_0&=&0,\quad
\delta_0\circ \dM-(\dM\otimes 1+1\otimes\dM)\circ\delta_1=0,\\
\label{eq:closed3}d_\ad\delta_0(x,y)&=&0,\quad
d_\ad\delta_0(x,h)+d_\ad\delta_1(x,h)=0.
\end{eqnarray}

\section{Strict Lie 2-bialgebras}\label{sect:strict-lie2-bi}
\emptycomment{
A {\bf strict Lie 2-bialgebra} is a Lie 2-bialgebra with
$c_3=l_3=0$. }In this section, we study strict Lie 2-bialgebras in a more classical
setting.

\subsection{Manin triple and matched pairs} \label{sect:manin-triple}

Similar to  Definition \ref{defi:manin-triple},  a  {\bf
Manin triple of strict Lie $2$-algebras}, which we denote by
$(\huaK;\huaG,\huaG')$,  consists of
\begin{itemize}
\item three strict Lie $2$-algebras $\huaK, \huaG, \huaG^\prime$, where  $\huaG$ and $\huaG^\prime$ are sub-Lie $2$-algebras of $\huaK$,
 and $\huaK=\huaG\oplus \huaG'$ as vector space complexes;

\item  a degree $1$ nondegenerate invariant symmetric bilinear form $S_\huaK$ on $\huaK$,  such that
$\huaG$ and $\huaG'$ are isotropic.
\end{itemize}


A {\bf homomorphism}  between two Manin triples
$(\huaK_1;\huaG,\huaG')$ and $(\huaK_2;\huaH,\huaH')$ is a
homomorphism
 $\phi:\frak \huaK_1\rightarrow \huaK_2$ of strict Lie $2$-algebras
 satisfying,
\begin{equation}
\phi (\huaG)\subset \huaH, \quad\phi(\huaG')\subset \huaH',\quad
S_{\huaK_1}(\alpha,\beta)=S_{\huaK_2}(\phi(\alpha),\phi(\beta)),\;\forall~
\alpha,\beta\in \frkk_1.
\end{equation}

Let $\huaG=({\frak g}_0,{\frak g}_{-1}, \dM, [\cdot,\cdot])$ be a
strict Lie 2-algebra and suppose that there is also a strict Lie
2-algebra structure on $\huaG^*=({\frak g}_{-1}^*,{\frak g}_0^*,
\dM^*, [\cdot,\cdot]^*)$. On the direct sum of complexes, ${\frak
g}_{-1}\oplus{\frak g}_0^*\stackrel{\dM+\dM^*}{\longrightarrow}{\frak
g}_0\oplus {\frak
g}_{-1}^*$, there is a natural degree $1$
nondegenerate symmetric bilinear form $S$ given by,
\begin{equation}\label{eq:pair}
S(x+h+x^*+h^*,y+k+y^*+k^*)=\langle x, y^*\rangle+\langle h,
k^*\rangle+ \langle x^*, y\rangle+\langle h^*, k\rangle.
\end{equation}
 We call \eqref{eq:pair} the {\bf standard
bilinear form} on $\huaG\oplus \huaG^*$. We can introduce a bracket
operation $[\cdot,\cdot]_{\huaG\oplus \huaG^*}$, such that $S$ is
invariant, as follows,
\begin{equation}\label{eq:bracket}
\left\{\begin{array}{ll}~[x+h^*,y+k^*]_{\huaG\oplus \huaG^*}
=[x,y]+[h^*,k^*]^*+\ad_0^*(x)(k^*)-\add_0^*(k^*)(x)+\add_0^*(h^*)(y)-\ad_0^*(y)(h^*),\\
~[x+h^*,k+y^*]_{\huaG\oplus \huaG^*}
=[x,k]+[h^*,y^*]^*+\ad_0^*(x)(y^*)-\add_1^*(y^*)(x)+\add_0^*(h^*)(k)-\ad_1^*(h)(k^*).
\end{array}\right.
\end{equation}
We call this the {\bf standard bracket operation} on $\huaG\oplus
\huaG^*$, where $\add^*=(\add_0^*,\add_1^*)$ is the coadjoint
representation of $\huaG^*$ on $\huaG$. If
 $\huaG\oplus \huaG^*=({\frak g}_0\oplus {\frak
g}_{-1}^*,{\frak g}_{-1}\oplus{\frak g}_0^*, \dM+\dM^*,
[\cdot,\cdot]_{\huaG\oplus \huaG^*})$ is a strict Lie 2-algebra (in
this case, $\huaG$ and $\huaG^*$ are sub-Lie 2-algebras naturally),
then we obtain a Manin triple $(\huaG\oplus \huaG^*;\huaG, \huaG^*)$
with respect to the standard bilinear form \eqref{eq:pair}, which we
call the {\bf standard Manin triple} of strict Lie 2-algebras.

\begin{pro} Any  Manin triple of strict Lie $2$-algebras
$(\huaK; \huaG, \huaG')$ with respect to a degree $1$ nondegenerate
invariant symmetric bilinear form $S_\huaK$ is isomorphic to a
standard Manin triple $(\huaG\oplus \huaG^*;\huaG, \huaG^*)$.
\end{pro}

\pf  The
nondegeneracy of  $S_\huaK$ implies that
$\g'$ is isomorphic to $\g^*$. The invariancy of  $S_\huaK$ further
tells us that the bracket operation must be given by
\eqref{eq:bracket}. \qed\vspace{3mm}

Now we consider how to define a strict Lie 2-algebra structure on
the direct sum $\huaG\oplus \huaG'$
 of two strict Lie 2-algebras $\huaG$ and $\huaG'$ such
that they are strict sub-Lie 2-algebras.
\begin{thm}\label{thm:directsum}
Let $\huaG$ and $\huaG^\prime$ be two strict Lie $2$-algebras,
$\mu=(\mu_0,\mu_1):\huaG\longrightarrow\End(\huaG^\prime)$, and
$\mu^\prime=(\mu_0',\mu_1'):\huaG^\prime\longrightarrow\End(\huaG)$
be representations of $\huaG$ and $\huaG^\prime$ on $\huaG^\prime$
and $\huaG$ respectively satisfying the following compatibility
conditions:
\begin{equation}\label{eq:c1}\left\{\begin{array}{rcl}
\mu_0'(x')[x,y]&=&[x,\mu_0'(x')y]+[\mu_0'(x')x, y]+\mu_0'(\mu_0(y)x')x-\mu_0'(\mu_0(x)x')y;\\
\mu_0(x)[x',y']'&=&[x',\mu_0(x)y']'+[\mu_0(x)x', y']'+\mu_0(\mu_0'(y')x)x'-\mu_0(\mu_0'(x')x)y';\\
 \mu_1'(h')[x,y]&=&[x,\mu_1'(h')y]+[\mu_1'(h')x, y]+\mu_1'(\mu_0(y)h')x-\mu_1'(\mu_0(x)h')y;\\
 \mu_1(h)[x',y']'&=&[x',\mu_1(h)y']'+[\mu_1(h)x', y']'+\mu_1(\mu_0'(y')h)x'-\mu_1(\mu_0'(x')h)y';\\
 \mu_0'(x')[x,h]&=&[x,\mu_0'(x')h]+[\mu_0'(x')x,h]+\mu_1'(\mu_1(h)x')x-\mu_0'(\mu_0(x)x')h;\\
\mu_0(x)[x',h']'&=&[x',\mu_0(x)h']'+[\mu_0(x)x',h']'+\mu_1(\mu_1(h')x)x'-\mu_0(\mu_0'(x')x)h'.\end{array}\right.
\end{equation}
Then there exists a strict Lie $2$-algebra
$(\frkg\oplus\frkg',\dM\oplus
\dM',[\cdot,\cdot]_{\huaG\oplus\huaG'})$, where
$[\cdot,\cdot]_{\huaG\oplus\huaG'}$ is given by
\begin{equation}\label{eq:sum1}
\left\{\begin{array}{l}
~[x+x',y+y']_{\huaG\oplus\huaG'}=[x,y]+\mu_0(x)(y')-\mu_0'(y')x+
\mu_0'(x')y-\mu_0(y)x'+
[x',y']';\\
~[x+x',h+h']_{\huaG\oplus\huaG'}=[x,h]+\mu_0(x)(h')-\mu_1'(h')(x)-\mu_1(h)x'+\mu_0'(x')(h)+[x',h']'.\end{array}\right.
\end{equation}

Conversely, given a strict Lie $2$-algebra $(\frak g\oplus\frak
g',\dM\oplus \dM',[\cdot,\cdot]_{\huaG\oplus\huaG'})$, in which
$\huag$ and $\huag'$  are strict sub-Lie $2$-algebras with respect
to the restricted brackets, there exist representations $\mu
:\huaG\longrightarrow\End(\huaG^\prime)$ and $\mu'
:\huaG^\prime\longrightarrow\End(\huaG)$ satisfying \eqref{eq:c1}
such that the bracket $[\cdot, \cdot]_{\huag\oplus
  \huag'}$ is given by \eqref{eq:sum1}.
\end{thm}

\pf We give a proof using the big bracket on
$Sym\big((\g\oplus\g')^*[-1]\otimes (\g\oplus\g') \big)$. Let
$(l_1,l_2)\in Sym(\g^*[-1])\otimes \g$ and $(l_1',l_2')\in
Sym({\g'}^*[-1])\otimes \g'$ be the Lie 2-algebra structures on
$\frkg$ and $\frkg'$ respectively.
  On the graded vector space $\g\oplus\g'=(\g_{-1}\oplus\g_{-1}')\oplus
 (\g_0\oplus\g_0')$,
 define  $(\overline{l_1},\overline{l_2})\in Sym((\g\oplus\g')^*[-1])\otimes (\g\oplus\g')$  by
 $\overline{l_1}=l_1+l_1'$ and
 $\overline{l_2}=l_2+l_2'+{l_2^\star}$, where ${l_2^\star}$ is given
 by
 \begin{eqnarray*}
  { l_2^\star }(x,y')=\mu_0(x)y'-\mu_0'(y')x,\quad
{l_2^\star} (x,h')=\mu_0(x)h'-\mu_1'(h')x, \quad{l_2^\star}
(x',h)=\mu_0'(x')h-\mu_1(h)x'.
 \end{eqnarray*} Clearly,
$\overline{l_2}$ is given by \eqref{eq:sum1}.
Equation \eqref{eq:c1} together with the fact that $\mu$ and $\mu'$
are strict representations implies that \( \pair{\overline{l_2}, \overline{l_2}}=0. \)
 Since $l_1\in\g_{-1}^*\otimes \g_0,~l_1'\in{\g_{-1}'}^*\otimes
 \g_0',~l_2\in\big((\wedge^2\g_0^*)\otimes
 \g_0\big)\oplus\big((\g_0^*\otimes \g_{-1}^*)\otimes \g_{-1}\big)$, and
$l_2'\in\big((\wedge^2{\g_0'}^*)\otimes
\g_0'\big)\oplus\big(({\g_0'}^*\otimes {\g_{-1}'}^*)\otimes
\g_{-1}'\big)$, it is obvious that
$$
\pair{l_1,l_1'}=0,\quad \pair{l_1,l_2'}=0,\quad
\pair{l_1',l_2}=0,\quad \pair{l_2,l_2'}=0.
$$
By the fact that both $\g$ and $\g'$ are strict Lie 2-algebras, we
have
$$
\pair{l_1,l_1}=0,\quad \pair{l_1,l_2}=0,\quad \pair{l_2,l_2}=0,\quad
\pair{l_1',l_1'}=0,\quad\pair{l_1',l_2'}=0,\quad \pair{l_2',l_2'}=0.
$$
Since $\mu$ and $\mu'$ are strict representations, we have
$\pair{l_1+l_1',
  l_2^\star}=0$.
   Thus, we have
$\pair{\overline{l_1}+\overline{l_2},\overline{l_1}+\overline{l_2}}=0$,
which implies that $(\overline{l_1},\overline{l_2})$ gives rise to a
Lie 2-algebra structure on $\g\oplus \g'$. The converse can be
 proved similarly. \qed

\begin{defi}\label{defi:matchpair}
  Let $\huaG=({\frak g}_0,{\frak g}_{-1}, \dM, [\cdot,\cdot])$ and
$\huaG^\prime=({\frak g}'_0,{\frak g}_{-1}', \dM', [\cdot,\cdot]')$
be two strict Lie $2$-algebras. Suppose that
$\mu=(\mu_0,\mu_1):\huaG\longrightarrow\End(\huaG^\prime)$ and
$\mu^\prime=(\mu_0',\mu_1'):\huaG^\prime\longrightarrow\End(\huaG)$
are representations of $\huaG$ and $\huaG^\prime$ on $\huaG^\prime$
and $\huaG$ respectively. We call them a {\bf matched pair} and
denote it by $(\huaG,\huaG^\prime;\mu,\mu^\prime)$ if
 \eqref{eq:c1} is satisfied.

A {\bf homomorphism} between two matched pairs
$(\huaG,\huaG^\prime;\mu,\mu^\prime)$ and
$(\frkh,\frkh^\prime;\nu,\nu^\prime)$ consists of strict Lie
$2$-algebra homomorphisms $f:\g\longrightarrow\frkh$ and
$f':\g'\longrightarrow\frkh'$ such that the following diagrams
commute:
$$
\xymatrix{
\frkg\times \g'\ar[r]^{\mu}\ar[d]^{f\times f'}&\g'\ar[d]^{f'}\\
\frkh\times\frkh'\ar[r]^{\nu}&\frkh',}  \qquad  \xymatrix{
\frkg'\times \g\ar[r]^{\mu'}\ar[d]^{f'\times f}&\g\ar[d]^{f}\\
\frkh'\times\frkh\ar[r]^{\nu'}&\frkh.}
$$

\end{defi}

\begin{pro}\label{pro:Manin-matchpair} Let $\huaG=({\frak g}_0,{\frak g}_{-1}, \dM, [\cdot,\cdot])$ and $\huaG^*=(\frak g_{-1}^*,\frak g_0^*,\dM^*, [\cdot,\cdot]^*)$
 be two strict Lie $2$-algebras.
 Then $(\huaG\oplus \huaG^*; \huaG, \huaG^*)$ is a
standard Manin triple if and only if
 $(\huaG, \huaG^*; \ad^*, \add^*)$ is a matched pair of strict Lie
 $2$-algebras. We call such a matched pair $(\huaG, \huaG^*;
 \ad^*, \add^*)$ a {\bf standard matched pair}.

 Furthermore, any
 homomorphism between standard Manin triples of strict Lie $2$-algebras induces a homomorphism between the corresponding matched pairs.
\end{pro}

\pf It is straightforward to see that the standard bilinear form
\eqref{eq:pair} is invariant under the standard bracket operation
\eqref{eq:bracket}. Furthermore, it is also not hard to deduce that
$\dM\oplus\dM^*$ is a graded derivation with respect to the standard
bracket operation \eqref{eq:bracket}. Thus $(\huaG\oplus \huaG^*;
\huaG, \huaG^*)$ is a standard Manin triple if and only if the
standard bracket operation \eqref{eq:bracket}  satisfies the Jacobi
identity. This is further equivalent to the fact that $(\huaG,
\huaG^*; \ad^*, \add^*)$ is a matched pair, by Theorem \ref{thm:directsum}.

Given a homomorphism between standard Manin triples $\phi:(\g\oplus
\g^*; \g, \g^*) \longrightarrow(\frkh\oplus\frkh^*; \frkh, \frkh^*)$,
denote by $\phi_\g$ the restriction $\phi|_\g$
and $\phi_{\g^*}$ the restriction $\phi|_{\g^*}$.  It is easy to see that both $\phi_\g
$ and $\phi_{\g^*}$ are strict homomorphisms of strict Lie
2-algebras. By the fact that $\phi$ gives a strict homomorphism of Lie 2-algebras $\g\oplus
\g^*\longrightarrow\frkh\oplus\frkh^*$, it is easy to see that the
diagrams in Definition \ref{defi:matchpair} commute. Thus,
$(\phi_\g,\phi_{\g^*})$ is a homomorphism between matched pairs.
\qed

\begin{rmk} \label{rmk:homo-mt-mp}
  Since a homomorphism $\phi:\g\oplus
\g^*\longrightarrow\frkh\oplus\frkh^*$ between standard Manin triples
  preserves the standard bilinear form, it forces $(\phi|_\frkg)^*$ to
  be the inverse of $\phi|_{\frkg^*}$.   Thus, a
  homomorphism of standard Manin triples must be an isomorphism. Therefore, the converse of the
  second part of the above proposition is not true: a homomorphism
  between two standard matched pairs can not induce a homomorphism of the corresponding standard Manin
triples in general.
\end{rmk}

For  linear maps $\delta_1:\frak g_{-1}\rightarrow \frak
g_{-1}\otimes \frak g_{-1}$ and $\delta_0:\frak g_0\rightarrow \frak
g_{-1}\otimes \frak g_0\oplus \frak g_0\otimes \frak g_{-1}$, define
$[\cdot,\cdot]^*:\frkg_{-1}^*\otimes\frkg_{-1}^*\longrightarrow\frkg_{-1}^*$,
$[\cdot,\cdot]^*:\frkg_{-1}^*\otimes\frkg_0^*\longrightarrow\frkg_{0}^*$
and
$[\cdot,\cdot]^*:\frkg_0^*\otimes\frkg_{-1}^*\longrightarrow\frkg_{0}^*$
by
\begin{eqnarray}
\label{eq:br5}\langle[h^*,k^*]^*,l\rangle\triangleq\langle
h^*\otimes k^*,\delta_1(l)\rangle,
\langle[h^*,x^*]^*,y\rangle\triangleq\langle h^*\otimes
x^*,\delta_0(y)\rangle, \langle[x^*,h^*]^*,y\rangle\triangleq\langle
 x^*\otimes h^*,\delta_0(y)\rangle.
\end{eqnarray}
\begin{rmk}\label{rk:degree}
We need to emphasize here that the above pairing $\langle \cdot,
\cdot \rangle$ is the usual pairing between
a vector space and its dual space, which is
different from the big bracket $\pair{\cdot, \cdot}$.

For a strict Lie $2$-bialgebra, we have $c_1=l_1$, $c_3=0$,  and
$c_2$ corresponds to  $[\cdot,\cdot ]^*$, and thus corresponds to $\delta_0$ and
$\delta_1$.  However, $c_2 \neq \delta_0+\delta_1$ (see Lemma
\ref{lem:c2d0d1}). This difference produces a slightly different
cohomological explanation of the compatibility relation  in Theorem
\ref{thm:delta-cocycle}.
\end{rmk}
\begin{lem}\label{lem:c2d0d1}
Let $\huaG=({\frak g}_0,{\frak g}_{-1}, \dM, [\cdot,\cdot])$ be a
strict Lie $2$-algebra, and  linear maps $\delta_1,\delta_0$
together with $\dM^*$  define a strict Lie $2$-algebra structure on
$\huaG^*$. Then the corresponding cobracket $c_2\in {\frak g}^*[-1]
\otimes Sym^2({\frak g}[-2])$ is a $4$-cocycle of $\frak g$ with
coefficients in $Sym (\frak g [-2])$ if and only if $(\delta_0,
\delta_1)$  is a $2$-cocycle of $\frak g$ with coefficients in $\frak
g \otimes \frak g$.
\end{lem}
 \pf If $c_2= m^* hk + y^* xn$ then $\delta_0=
y^*\otimes (x\otimes n-n\otimes x)$ and $\delta_1= m^*\otimes
(h\otimes k-k\otimes h)$. Then a routine calculation shows that
$\tilde{D}_{ad} c_2=0$ (see \eqref{eq:td}) if and only if $D
(\delta_0,\delta_1)=0$. \qed
\vspace{3mm}

Therefore, a
{\bf strict Lie $2$-bialgebra}  consists of a strict Lie $2$-algebra
$\huaG=({\frak g}_0,{\frak g}_{-1}, \dM, [\cdot,\cdot])$ and a
$2$-cocycle $(\delta_0,\delta_1)$, where $\delta_0,\delta_1$ define
a semidirect product Lie algebra structure $[\cdot,\cdot]^*$ on
$\frkg_{-1}^*\oplus\frkg_0^*$ via \eqref{eq:br5}. We denote a strict
Lie 2-bialgebra by $(\huaG;( \delta_0, \delta_1))$.
  A {\bf homomorphism} between strict Lie $2$-bialgebras $(\huaG;
(\delta_0, \delta_1))$ and $(\frkh;(\epsilon_0,\epsilon_1))$ is
defined to be a strict
homomorphism of strict Lie $2$-algebras
$f=(f_0,f_1):\frkg\longrightarrow \frkh$ such that $$(f_1\otimes
f_0+f_0\otimes f_1)\delta_0(x)=\epsilon_0(f_0(x)),\quad (f_1\otimes
f_1)\delta_1(h)=\epsilon_1(f_1(h)).$$

\emptycomment{
The following theorem is similar to Theorem
\ref{thm:c-cocycle}. However, for the purpose of
 potential generalization in the non-symmetric setting,  our cobrackets take values in
 the tensor algebra instead of the symmetric algebra as before.}

\begin{thm}\label{thm:delta-cocycle}
Let $\huaG=({\frak g}_0,{\frak g}_{-1}, \dM, [\cdot,\cdot])$ be a
strict Lie $2$-algebra. Suppose that the linear maps $\delta_1,
\delta_0$ define a semidirect product Lie algebra structure
$[\cdot,\cdot]^*$ on $\frkg_{-1}^*\oplus\frkg_0^*$ via
\eqref{eq:br5}. Then $\huaG^*=(\frak g_{-1}^*,\frak g_0^*,\dM^*,
[\cdot,\cdot]^*)$ is a strict Lie $2$-algebra such that $(\huaG,
\huaG^*; \ad^*, \add^*)$ is a matched pair of strict Lie
$2$-algebras if and only if $(\delta_0,\delta_1)$ is a $2$-cocycle
of $\huaG$ with coefficients in $\huaG\otimes \huaG$.
\end{thm}
\pf $(\delta_0,\delta_1)$ is a $2$-cocycle if and only if
\eqref{eq:closed1} and \eqref{eq:closed3} hold. It is
straightforward to see that \eqref{eq:closed1} is equivalent to the
compatibility of $\dM^*$ and $[\cdot,\cdot]^*$, which makes
$\huaG^*=(\frak g_{-1}^*,\frak g_0^*,\dM^*, [\cdot,\cdot]^*)$  a
strict Lie $2$-algebra. On the other hand, it is not hard to see
that \eqref{eq:closed3} is equivalent to \eqref{eq:c1}. Thus,
$(\huaG, \huaG^*; \ad^*, \add^*)$ is a matched pair of strict Lie
$2$-algebras. \qed

\emptycomment{The relation of Theorem \ref{thm:delta-cocycle} to Theorem
\ref{thm:c-cocycle} is implied by the following lemma:}

By Theorem \ref{thm:delta-cocycle}, we see immediately that
\begin{cor}
There is a one-to-one correspondence between standard  Manin triples
of strict Lie $2$-algebras, standard matched pairs,  and strict Lie
$2$-bialgebras.
\end{cor}

\begin{rmk} \label{rmk:homo-lie2bi-mp}

The preceding result shows that there exists a one-to-one
correspondence on the level of objects between the
category of strict Lie $2$-bialgebras,  standard
Manin triples, and  standard matched pairs. However, we conclude from
 Remark \ref{rmk:homo-mt-mp} that there is  no such
correspondence on  the level of morphisms. Therefore, these categories are not equivalent.
First, every homomorphism in the category of standard Manin
triples is an isomorphism. Thus this category is a groupoid,
making it different from the other two.
Moreover,  in general, given a homomorphism between strict Lie
$2$-bialgebras,
  there is no corresponding homomorphism between the corresponding
  matched pairs of strict Lie $2$-algebras. However, given an isomorphism between strict Lie
  $2$-bialgebras $f:(\huaG;
(\delta_0, \delta_1))\longrightarrow(\frkh;(\epsilon_0,\epsilon_1))$, it
is straightforward to see that
$(f,{f^*}^{-1}):(\g,\g^*;\ad^*,\add^*)\longrightarrow
(\frkh,\frkh^*;\ad^*,\add^*) $ is an isomorphism between matched
pairs.
\end{rmk}

\begin{rmk}[Lie bialgebras as strict Lie $2$-bialgebras] \label{rk:lie-bialgebra}
  For an arbitrary Lie algebra $\frkh$, it is obvious that
  $\frkh\stackrel{\Id}{\longrightarrow}\frkh$ is a strict Lie
  $2$-algebra. If $(\frkh,\frkh^*)$ is a Lie bialgebra, it
  is easy to show   that
  $(\frkh\stackrel{\Id}{\longrightarrow}\frkh,\frkh^*\stackrel{\Id}{\longrightarrow}\frkh^*;\ad^*,\add^*)$
  is a matched pair of strict Lie $2$-algebras. Thus,
  $(\frkh\stackrel{\Id}{\longrightarrow}\frkh;(\delta_0,\delta_1))$ is
  a strict Lie $2$-bialgebra, where $\delta_0,\delta_1$ are both given
  by the Lie algebra structures on $\frkh^*$. The above provides
  a way to  embed  the category of
  Lie bialgebras in the category of Lie $2$-bialgebras.
\end{rmk}

\subsection{Coboundary strict Lie 2-bialgebras }\label{sect:cybe}

In this section,   we consider coboundary strict Lie 2-bialgebras
$\big(\frkg; (\delta_0, \delta_1)\big)$,
i.e., the  cases where the $2$-cochain $(\delta_0,\delta_1)$ is an exact $2$-cocycle. For
any $1$-cochain $(r,\phi)\in (\huaG[-1]\otimes\huaG[-1])_{1}\oplus
\Hom(\frkg_{0},(\huaG[-1]\otimes\huaG[-1])_{0})$, we have (for
notations, see Section \ref{sec:strict-case}):
\begin{eqnarray*}
D(r,\phi)=\widehat{\dM^\otimes}r+d_\ad
r+\widehat{\dM^\otimes}\phi+\widehat{\dM}\phi+d_\ad\phi =\dM^\otimes
r+d_\ad r-\dM^\otimes \circ\phi-\phi\circ\dM+d_\ad\phi.
\end{eqnarray*}
Therefore, if $(\delta_0,\delta_1)=D(r,\phi)$ for some $1$-cochain
$(r,\phi)$, we must have
\begin{eqnarray}\label{eq:restriction}
  (\dM\otimes1-1\otimes\dM) r&=&0,\quad
 d_\ad\phi=0,\\
\label{eq:d0phi}\delta_0(x)&=&d_\ad r(x)+\widehat{\dM^\otimes}\phi(x)=[x\otimes 1+1\otimes x,r]-\dM^\otimes\circ\phi(x),\\
\label{eq:d1phi}\delta_1(h)&=&d_\ad
r(h)+\widehat{\dM}\phi(h)=[h\otimes 1+1\otimes h,r]-\phi(\dM h).
\end{eqnarray}

The following conclusion is straightforward and we omit the proof.

\begin{pro}\label{pro:br}
Let  $\delta_0$ and $\delta_1$ be given by ~\eqref{eq:d0phi} and
\eqref{eq:d1phi} for some $(r,\phi)$ satisfying
\eqref{eq:restriction}. If
$$[\alpha\otimes 1+1\otimes\alpha, r+ \sigma
(r)]=0,\quad\forall~\alpha\in \frak g_0\oplus \frak g_{-1},$$ where
$\sigma$ is the exchange operator which exchanges the two copies of $\huag[-1]$, and $ \phi^*(h^*\otimes
k^*)+\phi^*(k^*\otimes h^*)=0, $ then the bracket operations
$[\cdot,\cdot]^*$ defined by  \eqref{eq:br5} are skew-symmetric.
Under this assumption, we have
\begin{eqnarray*}
\label{eq:br1}~[h^*,k^*]^*=[h^*,k^*]_r-\dM^*\phi^*(h^*\otimes
k^*),\quad \label{eq:br2}~[h^*,x^*]^*=[h^*,x^*]_r-\phi^*(h^*\otimes
\dM^*x^*),
\end{eqnarray*}
where $[h^*,k^*]_r$ and $[h^*,x^*]_r$ are given by
\begin{eqnarray*}
  [h^*,k^*]_r\triangleq\ad_{r(h^*)}^*k^*-\ad_{r(k^*)}^*h^*,\quad\quad
~  [h^*,x^*]_r\triangleq\ad_{r(h^*)}^*x^*-\ad_{r(x^*)}^*h^*.
\end{eqnarray*}
Furthermore, $[\cdot,\cdot]^*:\wedge
^2\frkg_{-1}^*\longrightarrow\frkg_{-1}^*$ satisfies the Jacobi
identity if and only if
\begin{eqnarray}\label{eq:dualJacobi}
[[h^*,k^*]_r,l^*]_r-\dM^*\phi^*([h^*,k^*]_r,l^*)-[\dM^*\phi^*(h^*,k^*),l^*]_r+\dM^*\phi^*(\dM^*\phi^*(h^*,k^*),l^*)+c.p.=0,
\end{eqnarray}
and $[\cdot,\cdot]^*:\frkg_{-1}^*\wedge
\frkg_0^*\longrightarrow\frkg_{0}^*$ satisfies the Jacobi identity
if and only if
\begin{eqnarray}
\nonumber&&[[h^*,k^*]_r,x^*]_r+c.p.
-\phi^*([h^*,k^*]_r,\dM^*x^*)-\phi^*(\dM^*[k^*,x^*]_r,h^*)-\phi^*(\dM^*[x^*,h^*]_r,k^*)\\
\nonumber&&-[\dM^*\phi^*(h^*,k^*),x^*]_r-[\phi^*(k^*,\dM^*x^*),h^*]_r-[\phi^*(\dM^*x^*,h^*),k^*]_r\\
\label{eq:dualrep}&&+\phi^*(\dM^*\phi^*(h^*,k^*),\dM^*x^*)+\phi^*(\dM^*\phi^*(k^*,\dM^*x^*),h^*)+\phi^*(\dM^*\phi^*(\dM^*
x^*,h^*),k^*)=0.
\end{eqnarray}
\end{pro}

Since we require $d_\ad\phi=0$,  we choose $\phi=d_\ad\frkr$ for
some $\frkr\in\frkg_{-1}\otimes\frkg_{-1}$.

\begin{pro}\label{pro:exact}
 If $\phi=d_\ad\frkr$ for some
  $\frkr\in\frkg_{-1}\otimes\frkg_{-1}$, then we have
\begin{equation} \label{eq:exact}
\delta_0(x)=[x\otimes 1+1\otimes x,r-\dM^\otimes\frkr],\quad
\delta_1(h)=[h\otimes 1+1\otimes h,r-\dM^\otimes\frkr].
\end{equation}
\end{pro}
\pf Since $D^2=0$, we have
$
\widehat{\dM^\otimes}\circ d_\ad\frkr+d_\ad\circ
\widehat{\dM^\otimes}\frkr=0,
$
which implies that
$
d_\ad\circ \dM^\otimes \frkr=\dM^\otimes\circ d_\ad\frkr.
$
Thus, we have
$$
\delta_0(x)=d_\ad r(x)-d_\ad(\dM^\otimes
\frkr)(x)=d_\ad(r-\dM^\otimes \frkr)(x)=[x\otimes 1+1\otimes
x,r-\dM^\otimes \frkr].
$$
Also by $D^2=0$, we have
$
\widehat{\dM}(d_\ad\frkr)+d_\ad(\widehat{\dM^\otimes}\frkr)=0,
$
which implies that
$
d_\ad\frkr(\dM h)=d_\ad(\dM^\otimes\frkr)(h).
$
Thus, we have
$$
\delta_1(h)=d_\ad r(h)-d_\ad\frkr(\dM h)=d_\ad
(r-\dM^\otimes\frkr)(h)=[h\otimes 1+1\otimes h,r-\dM^\otimes\frkr].
$$
The proof is completed. \qed\vspace{3mm}


The following well-known result can be found in \cite{D}:

\begin{lem}\label{lem:rmatrix}
Let $\frkh$ be a Lie algebra and $\delta: \frak h\rightarrow \frak
h\otimes \frak h$ be a linear map. If there exists $r\in \frak
h\otimes \frak h$ such that
$$\delta (x)=[x\otimes 1+1\otimes x, r],\;\quad\forall ~x\in \frak h,$$
 then $\delta^*:\frak h^*\otimes \frak h^*\rightarrow \frak h^*$
defines a Lie algebra structure if and only if $r$ satisfies
\begin{itemize}
\item $[x\otimes 1+1\otimes x, r+ \sigma (r)]=0$;
\item $[x\otimes 1\otimes 1+1\otimes x\otimes 1+1\otimes 1\otimes x, [r_{12},r_{13}]+[r_{13},r_{23}]+[r_{12},r_{23}]]=0$,
\end{itemize}
for any $x\in \frak h$. The above equations make sense in the universal enveloping
algebra of ${\frak h}$ and for $r=\sum\limits_i a_i\otimes b_i$,
\begin{equation}r_{12}=\sum\limits_i a_i\otimes b_i\otimes 1;\quad r_{13}=\sum\limits_i
a_i\otimes1\otimes b_i;\quad r_{23}=\sum\limits_i 1\otimes
a_i\otimes b_i.\end{equation}
\end{lem}

\begin{rmk} In particular, the following equation
\begin{equation}[r_{12},r_{13}]+[r_{13},r_{23}]+[r_{12},r_{23}]=0
\end{equation}
is called the classical Yang-Baxter equation (CYBE) in the Lie
algebra $\frak h$. The  matrix corresponding to a solution $r$ of
the CYBE is called a classical $r$-matrix.
\end{rmk}

\begin{thm}\label{thm:coboundaryLie2bi}
Let $\huaG=({\frak g}_0,{\frak g}_{-1}, \dM, [\cdot,\cdot])$ be a
strict Lie $2$-algebra with two linear maps $\delta_0$ and
$\delta_{1}$ given by  \eqref{eq:exact} for $r\in \huaG_0 \otimes
\huaG_{-1} \oplus  \huaG_{-1} \otimes \huaG_0$ and $\frkr \in
\huaG_{-1} \otimes \huaG_{-1}$. Then $(\huaG; (\delta_0, \delta_1))$
is a strict Lie $2$-bialgebra if and only if  the following
conditions are satisfied:
\begin{itemize}
\item[\rm(a)] $[\alpha\otimes 1+1\otimes\alpha, R+ \sigma (R)]=0$,
\item[\rm(b)] $[\alpha\otimes 1\otimes 1+1\otimes \alpha\otimes 1+1\otimes 1\otimes \alpha,
[R_{12},R_{13}]+[R_{13},R_{23}]+[R_{12},R_{23}]]=0$,
\item[\rm(c)] $\dM^\otimes r=0$,
\end{itemize}
 for any $\alpha\in \frak g_0\oplus \frak g_{-1}$, where $R=r-\dM^\otimes \frkr=r-(\dM\otimes1+1\otimes\dM)\frkr$.
\end{thm}
\pf Since $(\delta_0,\delta_1) =D(r,d_\ad\phi)$ is an exact cocycle,
by Theorem \ref{thm:delta-cocycle}, we only need to show that
$[\cdot,\cdot]^*$ given by  \eqref{eq:br5} defines a semidirect
product Lie algebra structure on $\frkg_{-1}^*\oplus\frkg_0^*$. The
conclusion follows from Proposition \ref{pro:exact} and Lemma
\ref{lem:rmatrix}. \qed\vspace{3mm}

Inspired by this theorem, we  call the classical Yang-Baxter
equation for $R$ together with $\dM^\otimes r=0$ the {\em
$2$-graded classical Yang-Baxter
  equations} ($2$-graded  CYBE) for $r$ and $\frkr$.
We have seen that for coboundary strict Lie $2$-bialgebras, there
are more general $r$-matrices, which are certain pairs $(r,\phi)$.
However, without requiring that $\phi=d_\ad \frkr$, it is not easy
to write down the equations that they need to obey.


\section{Constructions of strict Lie 2-bialgebras } \label{sect:ep}

In this section, we give explicit examples of strict Lie
2-bialgebras by solving the 2-graded  CYBE. These solutions are
constructed from left-symmetric algebras and symplectic Lie
algebras. We consider the case where   both $\delta_0$ and $\delta_1$
are given by $r\in\frkg_0\otimes
\frkg_{-1}\oplus\frkg_{-1}\otimes\frkg_0$, i.e.,
\begin{eqnarray*}
\delta_0(x)=[x\otimes 1+1\otimes x,r],\quad \delta_1(h)=[h\otimes
1+1\otimes h,r].
\end{eqnarray*}
By Theorem \ref{thm:coboundaryLie2bi}, if $r$ is a solution of CYBE
in $\frkg_0\oplus\frkg_{-1}$ and satisfies $\dM^\otimes r=0$, then
$r$ gives a strict Lie 2-bialgebra structure on the strict Lie
2-algebra $\frkg$.

\subsection{Examples from left-symmetric algebras}
\begin{defi}  A {\bf left-symmetric algebra}, $A$, is a vector space equipped with a bilinear product $(x,y)\rightarrow x\circ y$
satisfying that for any $x,y,z\in A$, the associator
$(x,y,z)=(x\circ y)\circ z-x\circ(y\circ z)$ is symmetric in $x,y$,
i.e.,
$$(x,y,z)=(y,x,z),\;\;{\rm or}\;\;{\rm
equivalently,}\;\;(x\circ y)\circ z-x\circ(y\circ z)=(y\circ x)\circ
z-y\circ(x\circ z).$$
\end{defi}
 For any $x\in A$, let $L_x$
denote the left multiplication operator, i.e., $L_x(y)=x\circ y$ for
any $y\in A$. The following conclusion is known (\cite{leftsymm4}):
\begin{lem} \label{lem:sub-ad} Let $A$ be a left-symmetric algebra. The commutator
$ [x,y]=x\circ y-y\circ x$ defines a Lie algebra ${\frak g}(A)$,
which is called the {\bf sub-adjacent Lie algebra} of $A$ and $A$ is also
called a {\bf compatible left-symmetric algebra}  on the Lie
algebra ${\frak g}(A)$. Furthermore, $L:{\frak g}(A)\rightarrow
\gl(A)$ with $x\rightarrow L_x$ gives a representation of the Lie
algebra ${\frak g}(A)$, i.e.,
 $
[L_x,L_y]=L_{[x,y]}.$
\end{lem}

Let $L^*$ be the dual representation of the Lie algebra $\frkg(A)$
on $A^*$. Then there is a semidirect product Lie algebra structure
$[\cdot,\cdot]_s$ on $\frkg(A)\oplus A^*$, which is given by
\begin{equation}\label{eq:s}
[x+h,y+k]_s=[x,y]+L^*_xk-L^*_yh.
\end{equation}
We denote the corresponding Lie algebra by $\frak g(A) \ltimes_{L^*}
A^*$. Let $\{e_1,\cdots, e_n\}$ be a basis of $A$ and
$\{e_1^*,\cdots, e_n^*\}$ be the dual basis of $A^*$. We have

\begin{thm}{\rm \cite{Bai-Unifiedapproach}}
Let $A$ be a left-symmetric algebra. Then
\begin{equation}
r=\sum_{i=1}^n (e_i\otimes e_i^*-e_i^*\otimes
e_i)\label{eq:rrr}\end{equation} is a solution of CYBE in $\frak
g(A) \ltimes_{L^*} A^*$.
\end{thm}

This motivates us to construct (coboundary) strict Lie 2-bialgebras
from left-symmetric algebras by constructing an explicit solution of
the 2-graded  CYBE: Let $\frak g_0=\frak g(A), \frak g_{-1}=A^*$.
Then, $r\in \frak g_0\otimes \frak g_{-1}\oplus \frak g_{-1}\otimes
\frak g_0$ given by ~(\ref{eq:rrr})  is a solution of the CYBE in
$\frak g(A) \ltimes_{L^*} A^*$. Comparing to the 2-graded  CYBE,
we need to take care of  one more equation: $\dM^\otimes r=0$. This
leads to the following proposition, which follows by a
straightforward calculation:

\begin{pro}\label{pro:condition:d}
Let $A$ be a left-symmetric algebra. If  a linear map
$\dM:A^*\longrightarrow A$ satisfies
  \begin{itemize}
    \item[\rm(i)] $\dM[x,h]_s=[x,\dM h]_s,\quad[\dM h,k]_s=[h,\dM
    k]_s$;
    \item[\rm(ii)] $(\dM\otimes 1-1\otimes \dM)r=0,$
  \end{itemize}
  where $r$ is given by ~\eqref{eq:rrr}, then  $r$  defines a strict
  Lie
  $2$-bialgebra structure on the strict Lie $2$-algebra
  $(\frkg(A),A^*,\dM,[\cdot,\cdot]_s)$, where  $[\cdot,\cdot]_s$ is given by
\eqref{eq:s}.
\end{pro}

\begin{ex}{\rm
 It is obvious that $\dM=0$  satisfies the conditions in Proposition
\ref{pro:condition:d}, i.e., for any left-symmetric algebra $A$,
$(\frak g(A), A^*, 0, [\cdot,\cdot]_s)$ is a strict Lie $2$-algebra
and, thus, there is a strict Lie $2$-bialgebra induced by $r$ given
by \eqref{eq:rrr}.}
\end{ex}
In  general, assume that $ \dM(e_i^*)=\sum_{j=1}^n d_{ij}e_j,
~i=1,\cdots, n. $ Let $M(\dM)=(d_{ij})$ be the corresponding matrix.
It is obvious that  $(\dM\otimes 1-1\otimes \dM)r=0$ if and only if
$M(\dM)$ is skew-symmetric.

\begin{ex} {\rm
The 1-dimensional non-trivial left-symmetric algebra is isomorphic
to the field, that is, there is a basis $\{e\}$ satisfying $e\circ
e=e$. In this case, it is straightforward to show that $\dM$
satisfies Condition (i) in Proposition \ref{pro:condition:d} if and
only if $\dM=0$.}
\end{ex}

The classification of 2-dimensional complex left-symmetric algebras
was given in \cite{Bai:classification,classification}.

\begin{ex}{\rm
A  non-trivial 2-dimensional complex left-symmetric algebra with $\dM$ satisfying
Condition (i) in Proposition \ref{pro:condition:d} is isomorphic to
one of the followings (we only give the non-zero products):
\begin{enumerate}
\item[A1.] $e_1\circ e_1=e_1, e_2\circ e_2=e_2:$ $\dM=0$;
\item[A2.] $e_1\circ e_1=e_1:$ $M(\dM)=\left( \begin{matrix} 0& a\cr 0 &b\cr\end{matrix}\right), a,b\in \mathbb C$;
\item[A3.] $e_1\circ e_1=e_1, e_1\circ e_2=e_2\circ e_1=e_2:$ $\dM=0$;
\item[A4.] $e_1\circ e_1=e_2: $$M(\dM)=\left( \begin{matrix} 0& 0\cr a &b\cr\end{matrix}\right), a,b\in \mathbb C$;
\item[N1.] $e_2\circ e_1=-e_1, e_2\circ e_2=ke_2$, $k\ne 1$ or $e_2\circ e_2=e_1-e_2:$ $\dM=0$;
\item[N2.] $e_2\circ e_1=-e_1, e_2\circ e_2=e_2:$ $M(\dM)=\left( \begin{matrix} 0& 0\cr a &0\cr\end{matrix}\right), a\in \mathbb C$;
\item[N3.] $e_1\circ e_1=e_1, e_2\circ e_1=e_2:$  $M(\dM)=\left( \begin{matrix} 0& -a\cr a &0\cr\end{matrix}\right), a\in \mathbb C$;
\item[N4.] $e_1\circ e_2=le_1, e_2\circ e_1=(l-1)e_1, e_2\circ e_2=e_1+le_2$, $l\ne 0,1:$ $\dM=0$;
\item[N5.] $e_1\circ e_2=e_1, e_2\circ e_2=e_1+e_2:$  $M(\dM)=\left( \begin{matrix} a& -a\cr a &0\cr\end{matrix}\right), a\in \mathbb C$;
\item[N6.] $e_1\circ e_1=2e_1,e_1\circ e_2=e_2, e_2\circ e_2=e_1:$ $\dM=0$.
\end{enumerate}
Hence, type (N3) is the only case that $M(\dM)$ is skew-symmetric
and nonzero. Note that (N3) is
 associative. In addition, it is a Novikov algebra (a left-symmetric algebra with
commutative right multiplication operators), which corresponds to
the Poisson bracket of one-dimensional hydrodynamics \cite{BN}.
Moreover, it gives  the so-called ``conformal current type Lie
algebras'' in terms of the Balinksy-Novikov's affinization
\cite{PB}.}
\end{ex}

\begin{ex}\label{ep:r2}  {\rm
By the classification in the above example, it is straightforward to
see that $\mathbb R^2$ has the following left-symmetric algebra
structure
$
e_1\circ e_1=e_1, \quad e_2\circ e_1=e_2,
$
where $\{e_1,e_2\}$ is a basis of $\mathbb R^2$.  The corresponding
Lie algebra structure $(\frkg(\mathbb R^2),[\cdot,\cdot])$ is
\begin{equation}\label{eq:ex1}
[e_2,e_1]=e_2\circ e_1-e_1\circ e_2=e_2.
\end{equation}
The dual representation of $\frkg(\mathbb R^2)$ on $\frkg^*(\mathbb
R^2)$ is given by
\begin{equation}\label{eq:ex2}
L^*_{e_1}e_1^*=-e_1^*,\quad L^*_{e_1}e_2^*=0,\quad
L^*_{e_2}e_1^*=0,\quad L^*_{e_2}e_2^*=-e_1^*,
\end{equation}
where $\{e_1^*,~e_2^*\}$ is the dual basis. Any $M(\dM)=\left(
\begin{matrix} 0& -a\cr a &0\cr\end{matrix}\right), a\in \mathbb R$
satisfies Condition (i) in Proposition \ref{pro:condition:d}. In
particular, let $a=1$. Then we have $ \dM(e_1^*)=-e_2,\quad
\dM(e_2^*)=e_1. $ Thus  we obtain a strict Lie $2$-algebra
$(\frkg(\mathbb R^2),\frkg^*(\mathbb R^2),\dM,[\cdot,\cdot]_s)$,
where $[\cdot,\cdot]_s$ is determined by  \eqref{eq:s},
\eqref{eq:ex1} and \eqref{eq:ex2}. By Proposition
\ref{pro:condition:d}, $r$ given by \eqref{eq:rrr} defines a strict
Lie $2$-bialgebra structure on the strict Lie $2$-algebra
$(\frkg(\mathbb R^2),\frkg^*(\mathbb R^2),\dM,[\cdot,\cdot]_s)$.
More precisely, we have
\begin{eqnarray*}
  \delta_0(e_1)&=&[e_1\otimes 1+1\otimes e_1,r]=-r=e_1^*\otimes
e_1-e_1\otimes e_1^*+e_2^*\otimes e_2-e_2\otimes e_2^*,\\
\delta_0(e_2)&=&0,\quad \delta_1(e_1^*)=0,\quad
\delta_1(e_2^*)=e_1^*\otimes e_2-e_2^*\otimes e_1.
\end{eqnarray*}
The dual complex of $\frkg^*(\mathbb
R^2)\stackrel{\dM}{\longrightarrow}\frkg(\mathbb R^2)$ is
$\frkg^*(\mathbb
R^2)\stackrel{\dM^*=-\dM}{\longrightarrow}\frkg(\mathbb R^2)$. The
Lie $2$-algebra structure $[\cdot,\cdot]^*$ on the dual complex is
given by
\begin{eqnarray*}
 [e_1,e_2]^*=e_2,\quad
{ [e_1,e_1^*]}^*=e_1^*,\quad {[e_2,e_2^*]}^*=e_1^*.
\end{eqnarray*}
In fact, we have
\begin{eqnarray*}
 \langle[e_1,e_2]^*,e_2^*\rangle=\langle\delta_1(e_2^*),e_1\otimes
 e_2\rangle=1,\quad
\langle[e_1,e_2]^*,e_1^*\rangle=\langle\delta_1(e_1^*),e_1\otimes
 e_2\rangle=0,
\end{eqnarray*}
which implies that $[e_1,e_2]^*=e_2$. The  others can be obtained
similarly.}
\end{ex}

\subsection{Examples from symplectic Lie algebras}

In the sequel, we consider the case where $\dM$ is invertible. We
find that it has a close relationship with symplectic Lie algebras,
which leads to an unexpected construction of strict Lie
2-bialgebras.

Let $\dM:A^*\rightarrow A$ be an invertible linear map such that
$M(\dM)$ is skew-symmetric. Then,
\begin{equation}\label{eq:Bd}
B_\dM(x,y)=\langle \dM^{-1}(x), y\rangle
\end{equation}
is a skew-symmetric nondegenerate bilinear form on $A$. Let $(A,
\circ)$ be a left-symmetric algebra. A skew-symmetric bilinear form
$\omega$ on $(A,\circ)$ is called {\bf invariant} if $\omega$
satisfies
$$
\omega(x\circ y, z) + \omega(z\circ x, y) - \omega(x\circ z, y) = 0,
$$
or equivalently, $ \omega(x\circ y, z)=\omega([x,z],y). $

\begin{pro} Let $(A,\circ)$  be a left-symmetric algebra and $\dM:A^*\rightarrow A$ be an invertible linear map such that $M(\dM)$ is skew-symmetric.
Then $(\frak g(A), A^*, \dM, [\cdot,\cdot]_s)$ is a strict Lie
$2$-algebra if and only if $B_\dM$ is invariant. In this case, $r$
given by ~\eqref{eq:rrr} gives rise to a Lie $2$-bialgebra
structure.
\end{pro}

\pf $(\frak g(A), A^*, \dM,  [\cdot,\cdot]_s)$ is a strict Lie
2-algebra if and only if $\dM$ satisfies
\begin{eqnarray*}
\dM[x,h]_s=[x,\dM h]_s,\qquad [h,\dM k]_s=[\dM h, k]_s,
\end{eqnarray*}
for any $x\in A, h, k\in A^*$. Set $\dM h=y$, $\dM k=z$. Then
$\dM[x,h]_s=[x,\dM h]_s$ holds if and only if
\begin{eqnarray*}
B_\dM([x,y]_s,z)&=&B_\dM([x,\dM h]_s,z)=B_\dM(\dM[x,h]_s,z)=\langle
[x,h]_s, z\rangle =\langle h, -x\circ z\rangle\\\nonumber
&=&B_\dM(\dM h, -x\circ z) =-B_\dM(y, x\circ z)=B_\dM(x\circ z,y ),
\end{eqnarray*}
which is exactly the condition that $B_\dM$ is invariant.
Furthermore, $[h,\dM k]_s=[\dM h, k]_s$ holds if and only if
\begin{eqnarray*}
B_\dM(y, z\circ x)=\langle h, \dM k\circ x\rangle=\langle [h, \dM
k]_s, x\rangle= \langle [\dM h, k]_s, x\rangle=-\langle k, \dM
h\circ x\rangle =-B_\dM(z, y\circ x).
\end{eqnarray*}
If $B_\dM$ is invariant, then the following can be obtained:
$$B_\dM(y,z\circ x)=-B_\dM(z\circ x,y)=-B_\dM([z,y],x)=B_\dM([y,z],x)=B_\dM(y\circ x,z)=-B_\dM(z,y\circ x).$$
Hence, the conclusion holds. \qed\vspace{3mm}

Let $\frkh$ be a Lie algebra, recall that a skew-symmetric bilinear
form $\omega$ is called a  2-cocycle if $\omega$ satisfies
$$
\omega([x,y],z) + \omega([y,z],x) + \omega([z,x],y) = 0,\;\;
\forall~ x, y, z \in \frak \frkh .
$$
A {\bf symplectic Lie algebra} is a pair $(\frkh,\omega)$, where
$\frkh$ is a Lie algebra and $\omega$ is a nondegenerate 2-cocycle.
The following result is given in \cite{chu}:

\begin{pro}  Let $(\frkh,\omega)$ be a symplectic Lie algebra. Then there exists a
compatible left-symmetric algebra structure $``\circ "$ on $\frkh$
given by
\begin{equation}\label{eq:leftsymmfromsym}
\omega (x\circ y,z) = -\omega (y,[x,z]),\quad\forall~ x, y, z \in
\frkh.
\end{equation}
Moreover, the Lie algebra $\frkh$ is the sub-adjacent Lie algebra of
this left-symmetric algebra (see Lemma \ref{lem:sub-ad}).
\end{pro}

\begin{pro}
  If $\omega$ is an invariant skew-symmetric bilinear form on  a
  left-symmetric algebra $(A,\circ)$, then $\omega$ is a $2$-cocycle
  of the sub-adjacent Lie algebra $\frak g(A)$; Conversely, if
  $(\frkh,\omega)$ is a symplectic Lie algebra, then $\omega$
is invariant with respect to the compatible left-symmetric algebra
structure given by \eqref{eq:leftsymmfromsym}.
\end{pro}
\pf If $\omega$ is invariant on $(A,\circ)$, then we have
$$\omega(x\circ y, z) - \omega(y\circ x, z) = \omega ([x, y], z)=
\omega ([x, z], y) - \omega([y, z], x), \;\;\forall x,y,z\in A.$$ So $\omega$ is a 2-cocycle
of the sub-adjacent Lie algebra $\frak g (A)$. The second half part
is obvious. \qed\vspace{3mm}

 This shows that a left-symmetric algebra with a nondegenerate (skew-symmetric)
invariant bilinear form is equivalent to a symplectic Lie algebra.

Summarizing the content of this section, we have the following
result.

\begin{thm}
Let $A$ be a left-symmetric algebra and $\dM:A^*\rightarrow A$ be an
invertible linear map such that $M(\dM)$ is skew-symmetric. Then the
following conditions are equivalent:
\begin{enumerate}
\item $(\frak g(A), A^*, \dM, [\cdot,\cdot]_s)$ is a strict Lie $2$-algebra;
\item The bilinear form $B_\dM$ induced by $\dM$ through  \eqref{eq:Bd} is invariant on $A$;
\item The sub-adjacent Lie algebra $\frak g(A)$ is a symplectic Lie algebra with the symplectic form
$B_\dM$ induced by $\dM$ through  \eqref{eq:Bd}.
\end{enumerate}
\end{thm}

\begin{cor}
Let $(\frkh,\omega)$ be a symplectic Lie algebra. Denote by
$(A,\circ)$ the corresponding left-symmetric algebra. Then $(\frak
g(A)=\frkh, A^*, \dM, [\cdot,\cdot]_s)$ is a strict Lie $2$-algebra,
where $[\cdot,\cdot]_s$ is given by the semidirect product Lie
algebra structure on $\frak g(A) \ltimes_{L^*} A^*$, in which the
compatible left-symmetric algebra structure is given by
\eqref{eq:leftsymmfromsym}, and $\dM$ is given by  \eqref{eq:Bd}, where $B_d=\omega$.
Moreover, $r$ given by  \eqref{eq:rrr} defines a strict Lie
$2$-bialgebra structure on $(\frak g(A), A^*, \dM,
[\cdot,\cdot]_s)$.\label{constr}
\end{cor}

\begin{ex}\label{ep:symp} {\rm
The study of symplectic Lie algebras is fruitful.  In particular,
there is a bialgebra theory of left-symmetric algebras which is
equivalent to a special class of symplectic Lie algebras
 that can be decomposed into a direct sum of two Lagrangian subalgebras \cite{Bai-bialgebra}. In the simplest case, for any
 left-symmetric algebra $(A,\circ)$,  there is a natural
symplectic Lie algebra structure on $A\oplus A^*$ whose Lie algebra
structure is given by $\frak g(A)\ltimes_{L^*} A^*$ and the
symplectic form is given by
\begin{equation}
\omega_p(x+a^*,y+b^*)=\langle a^*, y\rangle-\langle x, b^*\rangle,
\quad\forall ~x,y\in A, a^*,b^*\in A^*.\end{equation} The compatible
left-symmetric algebra structure, which we denote by
$\overline{\circ}$, on this symplectic Lie algebra defined by
\eqref{eq:leftsymmfromsym} is given by
\begin{equation}\label{eq:hatA}x\overline{\circ}y=x\circ y, ~x\overline{\circ}a^*=\ad^*_x
a^*,~
a^*\overline{\circ}x=\ad^*_xa^*-L^*_xa^*,~a^*\overline{\circ}b^*=0,
\end{equation}
for any $ x,y\in A, a^*,b^*\in A^*.$ Set $\widehat A=A\oplus A^*$.
Let $\{e_1,\cdots,e_n\}$ be a basis of $A$ and
$\{e_1^*,\cdots,e_n^*\}$ be the dual basis on $A^*$. Let
$\{f_1,\cdots,f_n,f_1^*,\cdots,f_n^*\}$ be the corresponding dual
basis on $\widehat{A}^*$, i.e.,
$$
\langle f_i,e_j\rangle=\delta_{ij},\quad \langle
f_i,e_j^*\rangle=0,\quad \langle f_i^*,e_j\rangle=0,\quad \langle
f_i^*,e_j^*\rangle=\delta_{ij}.
$$
By the definition of the symplectic form $\omega_p$, we can deduce
that $\dM$, which is determined by \eqref{eq:Bd}, is given by
$\dM(f_i)=e_i^*$ and $\dM(f^*_i)=-e_i$, or in terms of matrix, we
have $ M(\dM)=\left( \begin{matrix} 0& I_{n\times n} \cr -I_{n\times
n} &0\cr\end{matrix}\right). $ By Corollary \ref{constr}, there is a
strict Lie 2-algebra $(\frak g(\widehat A), {\widehat A}^*, \dM,
[\cdot,\cdot]_s)$, where $[\cdot,\cdot]_s$ is given by the
semidirect product Lie algebra structures on $\frak g(\widehat A)
\ltimes_{L_{\widehat A}^*} {\widehat A}^*$ associated to the
left-symmetric algebra structure \eqref{eq:hatA} on $\widehat A$.
Thus, $r=\sum_i(e_i\otimes f_i+e_i^*\otimes f_i^*-f_i\otimes
e_i-f_i^*\otimes e_i^*)$ gives rise to a strict Lie 2-bialgebra
structure. Note that this construction holds for any left-symmetric
algebra without any constraint condition. Thus this can be regarded as a
construction of strict Lie 2-bialgebras from the ``twice double
spaces'' of left-symmetric algebras. }
\end{ex}

\def\cprime{$'$} \def\cprime{$'$} \def\cprime{$'$} \def\cprime{$'$}

\end{document}